\documentclass[aps,pra,twocolumn,groupedaddress,amsmath,amssymb,superscriptaddress,reprint,showpacs,showkeys]{revtex4-1}
\usepackage{graphicx}  
\usepackage{dcolumn}   
\usepackage{bm}        
\usepackage{color}
\usepackage{float}
\usepackage{times}
\usepackage{amsmath}
\usepackage{multirow}
\usepackage{relsize}
\usepackage{booktabs} 
\usepackage{amsfonts, amsmath, amsthm, amssymb} 
\usepackage{adjustbox}
\usepackage[colorlinks=true,citecolor=blue,linkcolor=blue,urlcolor=blue]{hyperref}
\usepackage{color}
\usepackage{amssymb}
\usepackage{subfigure} 

\begin{document}
\title{Role of quantum correlations in light-matter quantum heat engines}
\author{G. Alvarado Barrios}
\email[Gabriel Alvarado]{\quad gabriel.alvarado@usach.cl}
\affiliation{Departamento de F\'isica, Universidad de Santiago de Chile (USACH), 
Avenida Ecuador 3493, 9170124, Santiago, Chile}

\author{F. Albarr\'an-Arriagada}
\affiliation{Departamento de F\'isica, Universidad de Santiago de Chile (USACH), 
Avenida Ecuador 3493, 9170124, Santiago, Chile}

\author{F. A. C{\'a}rdenas-L{\'o}pez}
\affiliation{Departamento de F\'isica, Universidad de Santiago de Chile (USACH), 
Avenida Ecuador 3493, 9170124, Santiago, Chile}

\affiliation{Center for the Development of Nanoscience and Nanotechnology 9170124, Estaci\'on Central, Santiago, Chile}

\author{G. Romero}
\affiliation{Departamento de F\'isica, Universidad de Santiago de Chile (USACH), 
	Avenida Ecuador 3493, 9170124, Santiago, Chile}

\author{J. C. Retamal}
\email[Juan Carlos Retamal~~~~]{\quad juan.retamal@usach.cl}
\affiliation{Departamento de F\'isica, Universidad de Santiago de Chile (USACH), 
Avenida Ecuador 3493, 9170124, Santiago, Chile}

\affiliation{Center for the Development of Nanoscience and Nanotechnology 9170124, Estaci\'on Central, Santiago, Chile}

\begin{abstract}
We study a quantum Otto engine embedding a working substance composed by a two-level system interacting with a harmonic mode. The physical properties of the substance are described by a generalized quantum Rabi model arising in superconducting circuits realizations. We show that light-matter quantum correlations reduction during the hot bath stage and compression stage act as a resource for enhanced work extraction and efficiency respectively. Also, we demonstrate that the anharmonic spectrum of the working subtance has a direct impact on the transition from heat engine into refrigerator as the light-matter coupling is increased. These results shed light on the search for optimal conditions in the performance of quantum heat engines.
\end{abstract}
\date{\today}
\maketitle
\section{Introduction}
Quantum heat engines (QHEs) \cite{scovil1959three} differ from classical heat engines in that they contain a quantum coherent working body, and the bounds imposed by quantum mechanics on their performance is a fundamental issue in quantum thermodynamics. They are characterized by three attributes: the working medium, the cycle of operation, and the dynamics of the governed cycle. Amongst the quantum cycles in which the engine may operate we have the Carnot cycle and the Otto cycle, with the quantum Otto cycle being the most widely studied \cite{feldmann1996heat,feldmann2004characteristics,rezek2006irreversible, Henrich2007,quan2007quantum,he2009performance,thomas2016performance}. With exception of the quantum Carnot cycle, the performance of the quantum cycle depends on the properties of the working substance, leading to several quantum heat engines proposals, namely, spin 1/2 and three-level systems \cite{He2002,feldmann2004characteristics,Geva1996,li2007quantum}, harmonic oscillators \cite{bender2002entropy,Kosloff2017harmonic}, and hybrid light-matter systems \cite{Hardal2015,song2016one}. In particular, for light-matter systems \cite{Hardal2015} the performance of the engine has been studied in different coupling regimes, exploring the relationship between  quantum coherence and correlations for enhancing work extraction. \par
A fundamental question whithin the study of QHEs with a working substance consisting of coupled quantum systems, is the role played by quantum correlations among the constituents on the performance of quantum heat machines. For an appropriately strong coupling it is possible to have non-zero quantum correlations even when the global system is in thermal equilibrium with some reservoir at a given temperature. A first insight into this problem has been given by Zhang et. al. \cite{Zhang2007} considering two interacting spins in a magnetic field linking the performance of the quantum Otto cycle to the entanglement present in the thermal equilibrium state. Additional efforts to understand this issue has been carried out by changing or generalizing the working substance \cite{Zhang2008,Wang2009,dillenschneider2009,Altintas2014,Hardal2015}. The main thing we have learned is that the effect of quantum correlations on the performance of quantum heat machines are model dependent.\par 
 Among the important things to dilucidate when chosing a working substance are: what is the effect of having the energy spectrum being degenerate or non-degenerate, the effect of the anharmonicity of the spectrum and the capability of the system for embedding quantum correlations. These are the main issues we address in the present work.\par
 We study a quantum Otto engine with a working substance composed by a two-level system interacting with a harmonic mode described by a generalized quantum Rabi model. Such physical device finds its realization in superconducting circuits \cite{niemczyk2010circuit,FornDiaz2010BSS,Bourassa2009USC,yoshihara2017DSC,forn2017ultrastrong,Yoshihara2017characteristic} where the interaction can be engineered to have access to different coupling regimes and different anharmonicities. In particular, we will study how the anharmonic spectrum of the quantum Rabi model \cite{Rabi1937,Braak2011} explains the change in operation regime as a function of the light-matter coupling. Afterwards, by considering the generalized quantum Rabi model we delve into the relation between the light-matter quantum correlations and the extractable work and efficiency of the engine. We show that contrary to previous works, the difference in quantum correlations between the thermal equilibrium states will not be indicative of the extractable work, instead, it is the quantum correlations reduction in the hot bath stage what can be interpreted as enhancing positive work. \par 
 This article is organized as follows, in section \ref{model} we describe the working substance of our QHE, in section \ref{QOE} we describe the quantum Otto cycle in which our QHE operates. In section \ref{Anharmonicity} we show the influence of the anharmonicity of the energy spectrum of the working substance on the operation regime of the QHE. In section \ref{Qcorrelations} and \ref{Qcorrelations_eff} we show the relation between the quantum correlation reduction during a thermodynamic process of the cycle and the harvested work and engine efficiency respectively. In section \ref{Optional} we consider an adiabatic process which simplifies a physical implementation of our QHE and study the quantum correlations and harvested work. Finally in section \ref{conclusions} we present our conclusions.\par 
\section{The model} \label{model}
Let us consider a QHE embedding a working substance consisting of a single cavity mode interacting with a two-level system which is described by the generalized quantum Rabi model \cite{niemczyk2010circuit,FornDiaz2010BSS,Bourassa2009USC,yoshihara2017DSC,forn2017ultrastrong,Yoshihara2017characteristic} : 
\begin{eqnarray}
\label{QRS} \nonumber
H = &&\hbar\omega_{\rm{cav}}a^{\dag}a + \frac{\hbar\omega_{q}}{2}\sigma^{z}\\  
  &+& \hbar g\big(\cos(\theta)\sigma^{x}+ \sin(\theta)\sigma^{z}\big)(a^{\dag} + a).
\end{eqnarray}
Here, $a(a^{\dag})$ is the annihilation (creation) bosonic operator for the field mode. The operators $\sigma^{z}$ and $\sigma^{x}$ stand for Pauli matrices describing the two-level system. Also, $\omega_{c}$, $\omega_{q}$,  $g$, and $\theta$, are the cavity frequency, qubit frequency, qubit-cavity coupling strength, and mixing angle respectively. In this work we will consider the resonance case for the qubit and cavity frequency $\omega_{\rm cav} = \omega_{q} = \omega$.\par
\begin{figure}[t!] 
	\centering
	\includegraphics[width= 0.99 \linewidth]{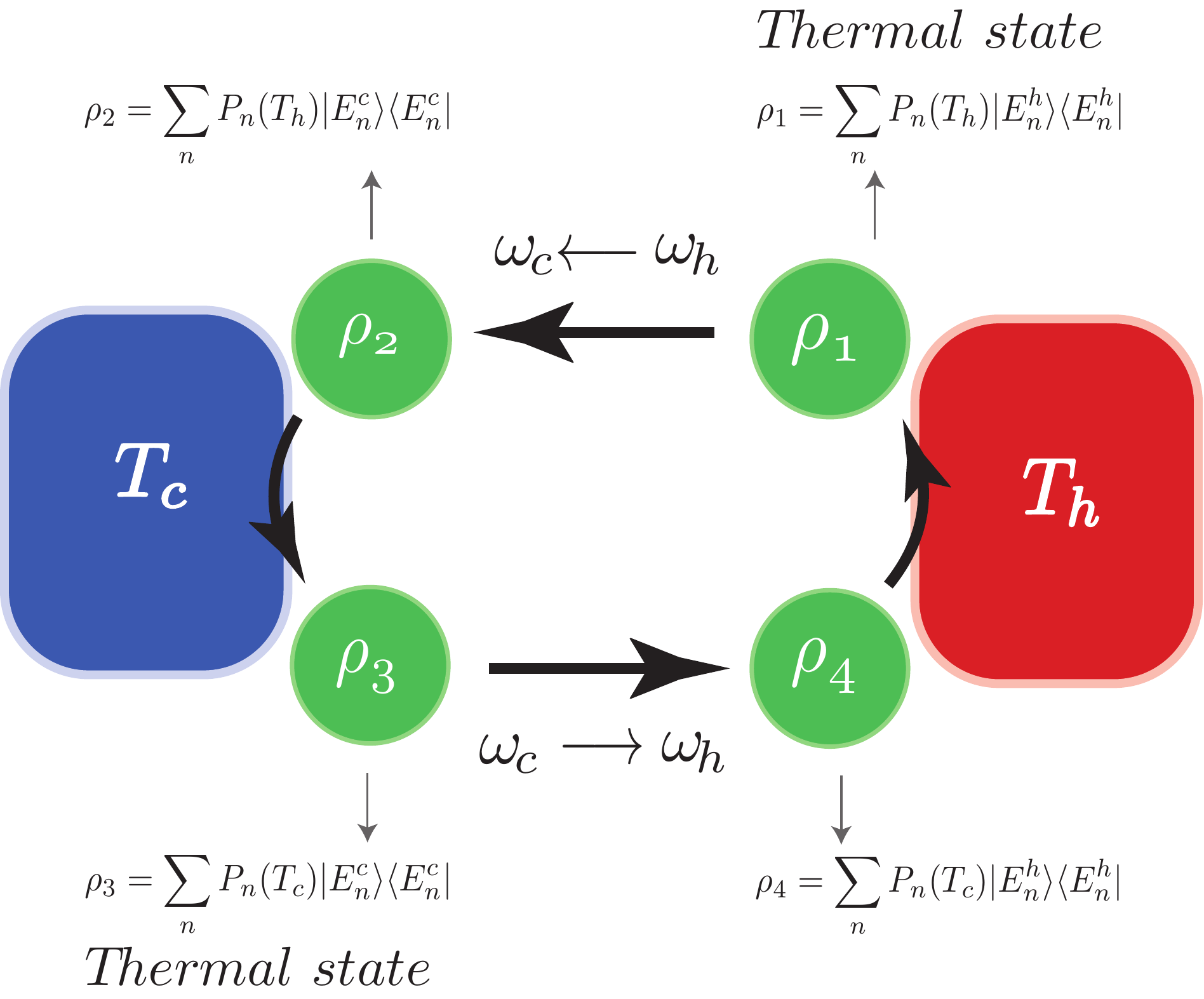}
	\caption{Diagram of the quantum Otto cycle utilized in this work, we can see the states at the different stages of the cycle. Where $T_{c}$ and $T_{h}$ are the temperatures of the cold and hot reservoir, respectively.}
	\label{fig:Ottocycle}
\end{figure}
It is worthwhile to note that for $\theta = 0$, this system corresponds to the quantum Rabi model \cite{Rabi1937,Braak2011} which has received increasing attention in recent years \cite{Nataf2011,Romero2012ultrafast,kyaw2015scalable,Kyaw2015QEC,joshi2017qubit}. The ratio of the coupling strength to the resonator frequency $g/\omega$ separates the behavior of the system into different regimes \cite{Wolf2013,Rossatto2016}. In the strong coupling regime, where the coupling strength is much larger than any decoherence or dephasing rate in the system, and for values $g/\omega \lesssim 10^{-2}$ one can perform the rotating wave approximation (RWA) and the system can be described by the Jaynes-Cummings model \cite{JC1963}. As the ratio $g/\omega$ is increased from the strong coupling regime there is a breakdown of the RWA and the system is described by the quantum Rabi model. We distinguish two main regimes for the later model, the ultra-strong coupling regime (USC) \cite{niemczyk2010circuit,FornDiaz2010BSS,Bourassa2009USC} where the coupling strength is comparable to the resonator frequency $g \lesssim \omega$ and the deep-strong coupling regime (DSC) \cite{yoshihara2017DSC,Casanova2010DSC} where the interaction parameter is greater than the relevant frequencies $g \gtrsim \omega$. In both regimes the eigenstates of the model correspond to highly correlated states of the  qubit and field mode. In addition the energy spectrum exhibits high anharmonicity as a function of the coupling strength. As we will show later, these properties of the model will explain the transition in operation regime of the quantum heat engine. \par
\section{Quantum Otto cycle}  \label{QOE}
In what follows we describe the thermodynamic cycle employed in our QHE. In this work we study a QHE operating under a quantum Otto cycle involving the working substance interacting with a cold and a hot reservoirs through four stages. We will consider that the coupling strength $g$ and the mixing angle $\theta$ will be kept constant throughout the cycle. The mode frequency will be changed from $\omega_{h}$ to $\omega_{c}$ to account for the interaction with the hot and cold reservoir respectively. The thermodynamic cycle used in this work is schematically presented in Fig.\ref{fig:Ottocycle}:\\
\begin{enumerate}
	\item Stage 1: Quantum isochoric process (hot bath stage). The system, with frequency $\omega = \omega_{h}$ and Hamiltonian $H_h$ is brought into contact with a hot thermal reservoir at temperature $T_h$ until it reaches thermal equilibrium. At the end of this process the state becomes $\rho_{1} = \sum_{n} P_{n}(T_{h})|E_{n}^{h}\rangle \langle E_{n}^{h}|$, where \{$|E_{n}^{h}\rangle$\} are the eigenstates of Hamiltonian $H_h$ and the corresponding thermal populations $P_{n}(T_{h})$ are given by the Boltzmann distribution for temperature $T_h$. It is noteworthy that this process will have a specific $g/\omega_{h}$ ratio for a given value of $g$, because we are fixing the coupling strength $g$ independently from the resonator frequency. During this process only the populations change while the energy level structure remains invariant.
	\item Stage 2: Quantum adiabatic (expansion) process. The system is isolated from the hot reservoir, and its frequency is changed from $\omega_{h}$ to $\omega_{c}$, with $\omega_{h}>\omega_{c}$, sufficiently slow as to satisfy the quantum adiabatic theorem such that the populations remain constant throughout the process.
	During this process only the energy level structure changes. At the end of this process the Hamiltonian is $H_c$ and the state of the system is $\rho_{2} = \sum_{n} P_{n}(T_{h})|E_{n}^{c}\rangle \langle E_{n}^{c}|$, where the thermal populations are the same as in $\rho_{1}$ but the energy eigenstates are those of $H_c$, therefore the system is no longer in a thermal state.
	\item Stage 3: Quantum isochoric process (cold bath stage). The working medium with $\omega = \omega_{c}$ and Hamiltonian $H_c$ is brought into contact with a cold thermal reservoir at temperature $T_c$ until it reaches thermal equilibrium. The state of the system at the end of this process is given by $\rho_{3} = \sum_{n} P_{n}(T_{c})|E_{n}^{c}\rangle \langle E_{n}^{c}|$, where \{$|E_{n}^{c}\rangle$\} are the energy eigenstates of $H_c$ and $P_{n}(T_{c})$ are the corresponding thermal populations at temperature $T_c$. Since the resonator frequency has changed to $\omega_{c}$ due to the adiabatic process, in this stage the ratio $g/\omega_{c}$ for a given value of $g$ is different than in stage 1 as a consequence of the adiabatic process.
	\item Stage 4: Quantum adiabatic (compression) process. The system is isolated from the cold reservoir, and its frequency is changed back from $\omega_{c}$ to $\omega_{h}$. During this process the populations remain unchanged while the energy level structure returns to its configuration in Stage 1. At the end of the process the Hamiltonian is $H_h$ and the state of the system is given by $\rho_{4} = \sum_{n} P_{n}(T_{c})|E_{n}^{h}\rangle \langle E_{n}^{h}|$, that is, the thermal populations are the same as in state $\rho_{3}$, but the energy eigenstates are those of $H_h$, thus, this is not a thermal state. \par The hot Hamiltonian $H_{h}$ and the cold Hamiltonian $H_{c}$ differ only by the frequency of the resonator, either $\omega_{h}$ or $\omega_{c}$. 
\end{enumerate}
We follow the conceptual frame as developed in \cite{PhysRevLett.93.140403} to define the heat transfered and work performed in a quantum thermodynamics. Let us consider the expectation value of the measured energy of a quantum system with discrete energy levels as given by $U=\langle E \rangle  = \sum_{n} p_{n}E_{n}$ where $E_{n}$ are the energy levels and $p_{n}$ are the corresponding occupation probabilities. Denote by $dU = \sum_{n}(p_{n}dE_{n} + E_{n}dp_{n})$ as the infinitesimal change of the energy, from which we can obtain the following identifications for infinitesimal heat transferred $dQ$ and work done $dW$ 
\begin{equation}
dQ := \sum_{n} E_{n} dp_{n}, \quad dW:= \sum_{n} p_{n} dE_{n}.
\end{equation}
 \begin{figure}[t!] 
	\centering
	\includegraphics[width= 1 \linewidth]{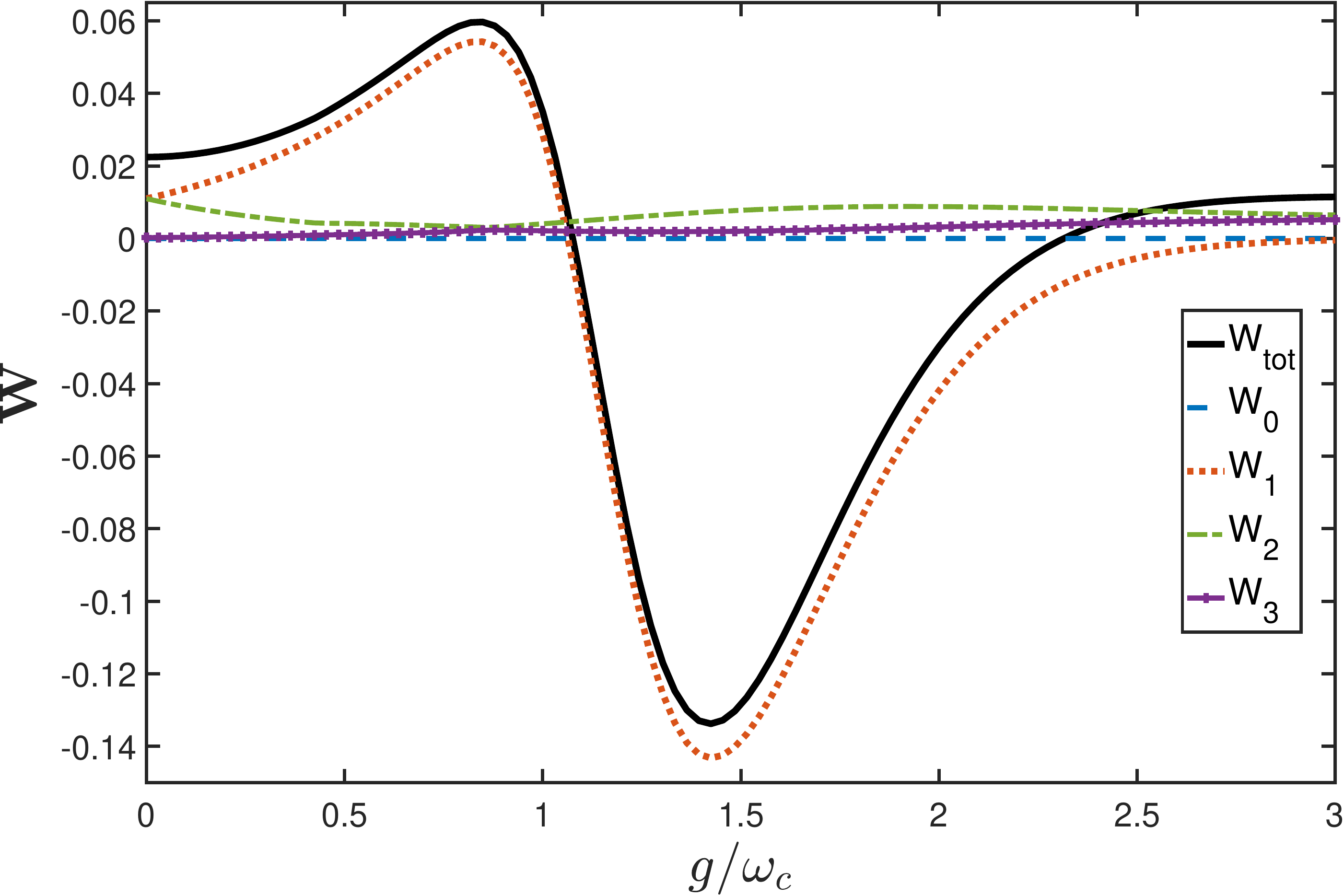}
	\caption{Work output of the quantum heat engine as a function of the coupling parameter $g/\omega_{c}$ for $\theta = 0$, with $\omega_{h} = 2\omega$ and $\omega_{c} = \omega$. We have used temperatures $T_{c} = 19$ mK and $T_{h} = 9T_{c}$.}
	\label{fig:work1}
\end{figure}
Here the heat transfer is related to the population change $dp_{n}$ with fixed energy level structure while the work done is related to the change in the energy levels $dE_{n}$ with fixed populations. The net work done in a single cycle can be obtained from $W = Q_h + Q_c$, where\\ 
 \begin{figure}[t!] 
	\centering
	\begin{subfigure}
		\centering
		\includegraphics[width=1\linewidth]{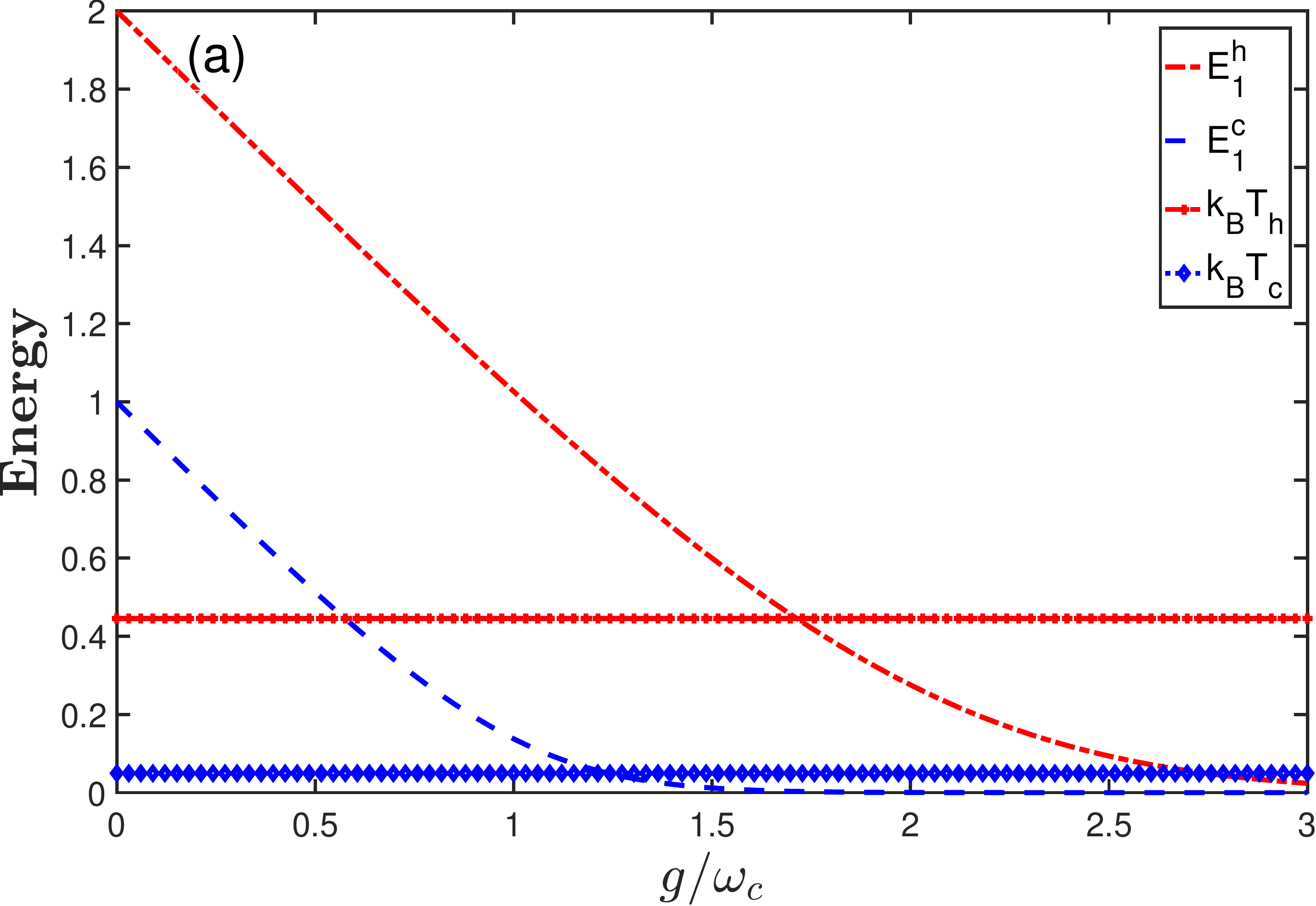}
		\label{fig:difEP}
	\end{subfigure}%
	\begin{subfigure}
		\centering
		\includegraphics[width=1\linewidth]{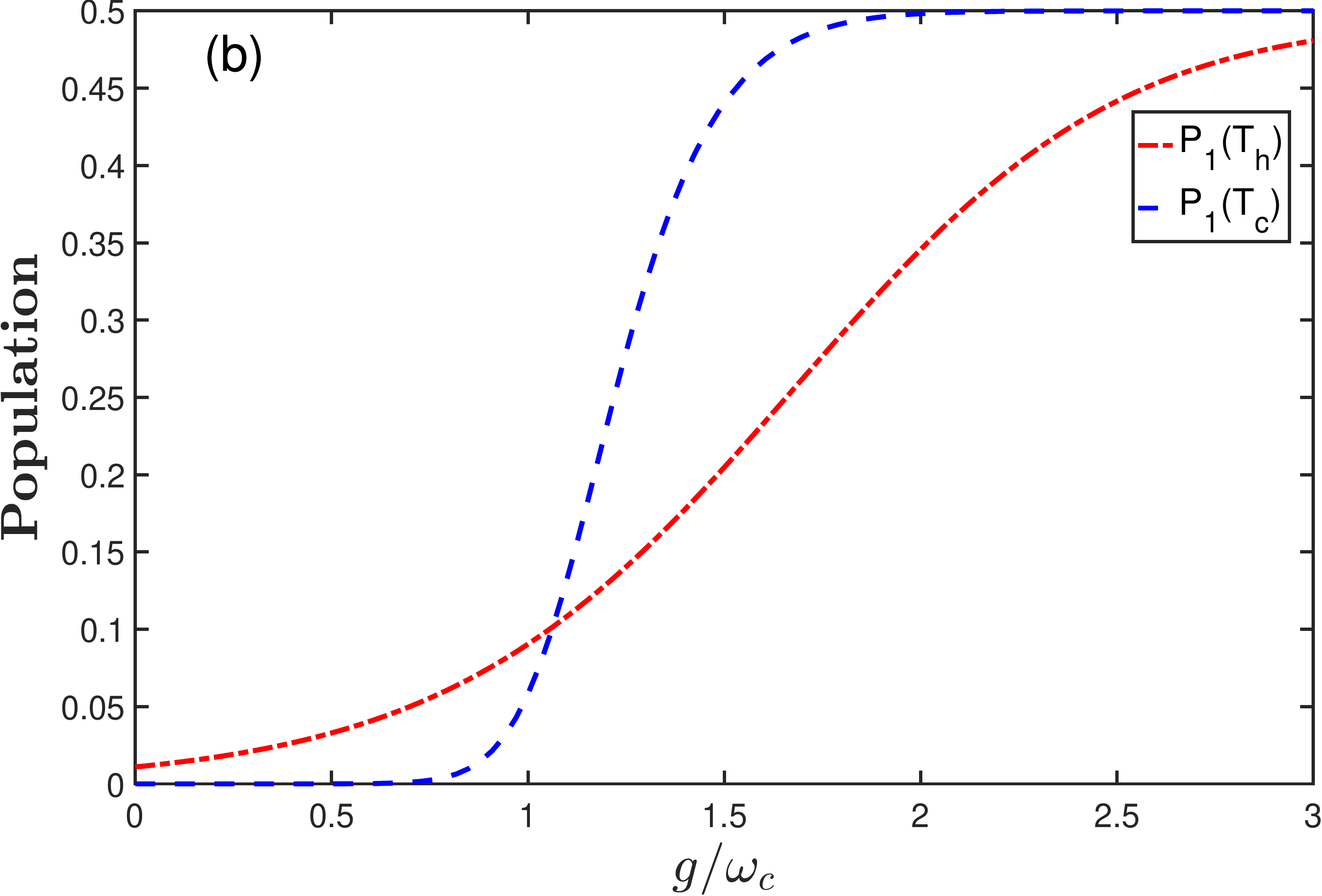}
		\label{fig:difP}
	\end{subfigure}%
	\caption{(a) Energy of the first excited state of the hot and cold Hamiltonian, and the thermal energy of the hot and cold reservoir. (b) Thermal population difference of the first excited state corresponding to the hot and cold Hamiltonians. We have used $\omega_{h} = 2\omega$ , $\omega_{c} = \omega$ with reservoir temperatures $T_{c} = 19$ mK and $T_{h} = 9T_{c}$.}
	\label{fig:3}
\end{figure}
\begin{eqnarray}
Q_{h} = \sum_{n} E_{n}^{h}\big(P_{n}(T_{h}) - P_{n}(T_{c})\big),\\
Q_{c} = \sum_{n} E_{n}^{c}\big(P_{n}(T_{c}) - P_{n}(T_{h})\big),\\
W = \sum_{n} \big( E_{n}^{h} - E_{n}^{c} \big)\big(P_{n}(T_{h}) - P_{n}(T_{c})\big).
\label{Work}
\end{eqnarray}
As can be seen in Eq.(\ref{Work}) the work performed by the engine receives a contribution per each energy level, but this contribution will only be relevant if the energy state is sufficiently thermally populated. Let us express the terms of Eq.(\ref{Work}) in the following way
\begin{eqnarray}
W =\sum_{n} W_{n},
\end{eqnarray}
where
\begin{eqnarray}
 W_{n} = \big( E_{n}^{h} - E_{n}^{c} \big)\big(P_{n}(T_{h}) - P_{n}(T_{c})\big).
\label{Work2}
\end{eqnarray}
Here we call $W_{n}$ the work contribution of the $n$-th energy level. Finally, the efficiency of the QHE is defined as the ratio between the extractable work and the heat that enters the system: 
\begin{eqnarray}
\eta = \frac{W}{Q_{h}}.
\end{eqnarray}
\begin{figure}[t!] 
	\centering
	\begin{subfigure}
		\centering
		\includegraphics[width=1 \linewidth]{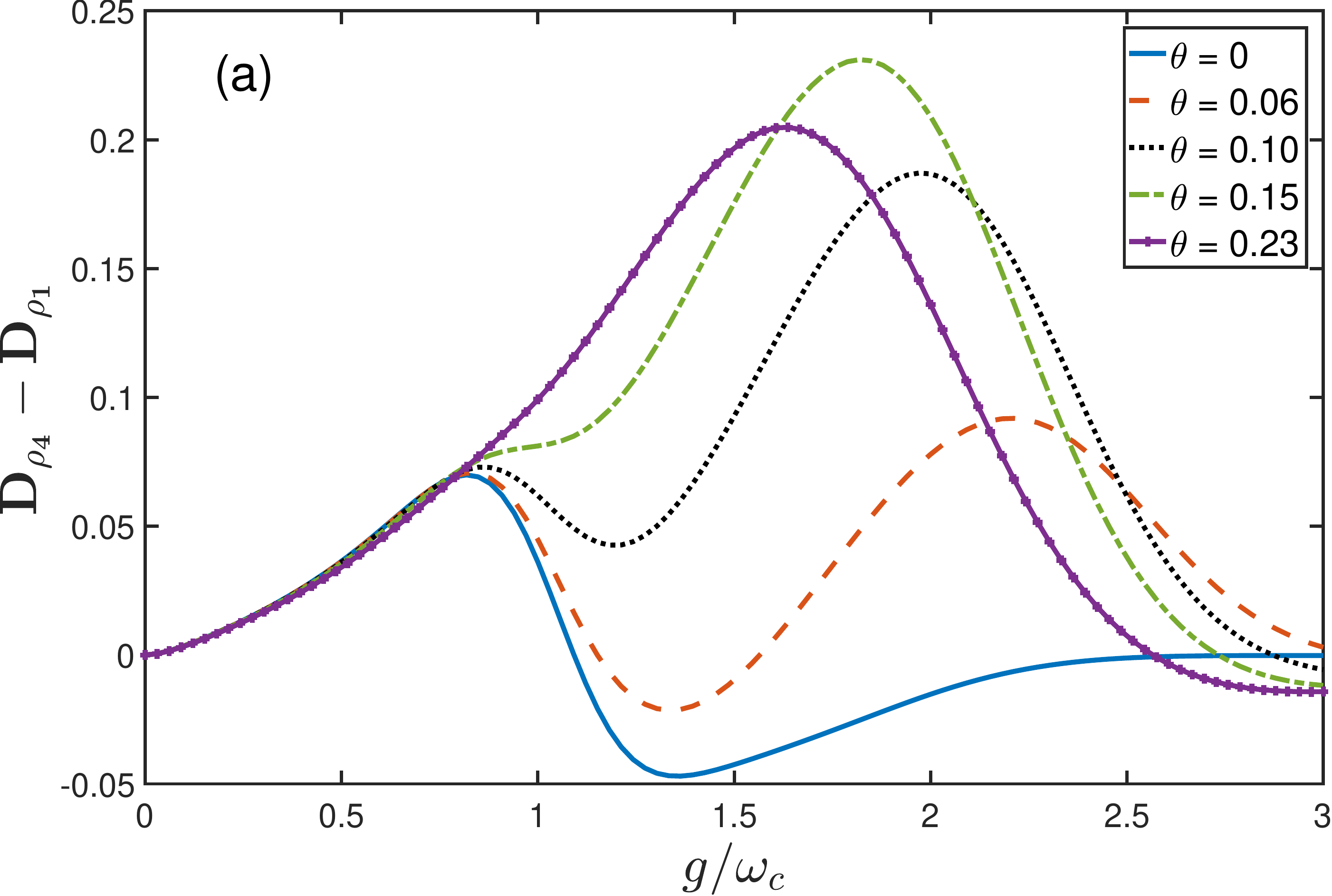}
	\end{subfigure}%
	\begin{subfigure}
		\centering
		\includegraphics[width=1 \linewidth]{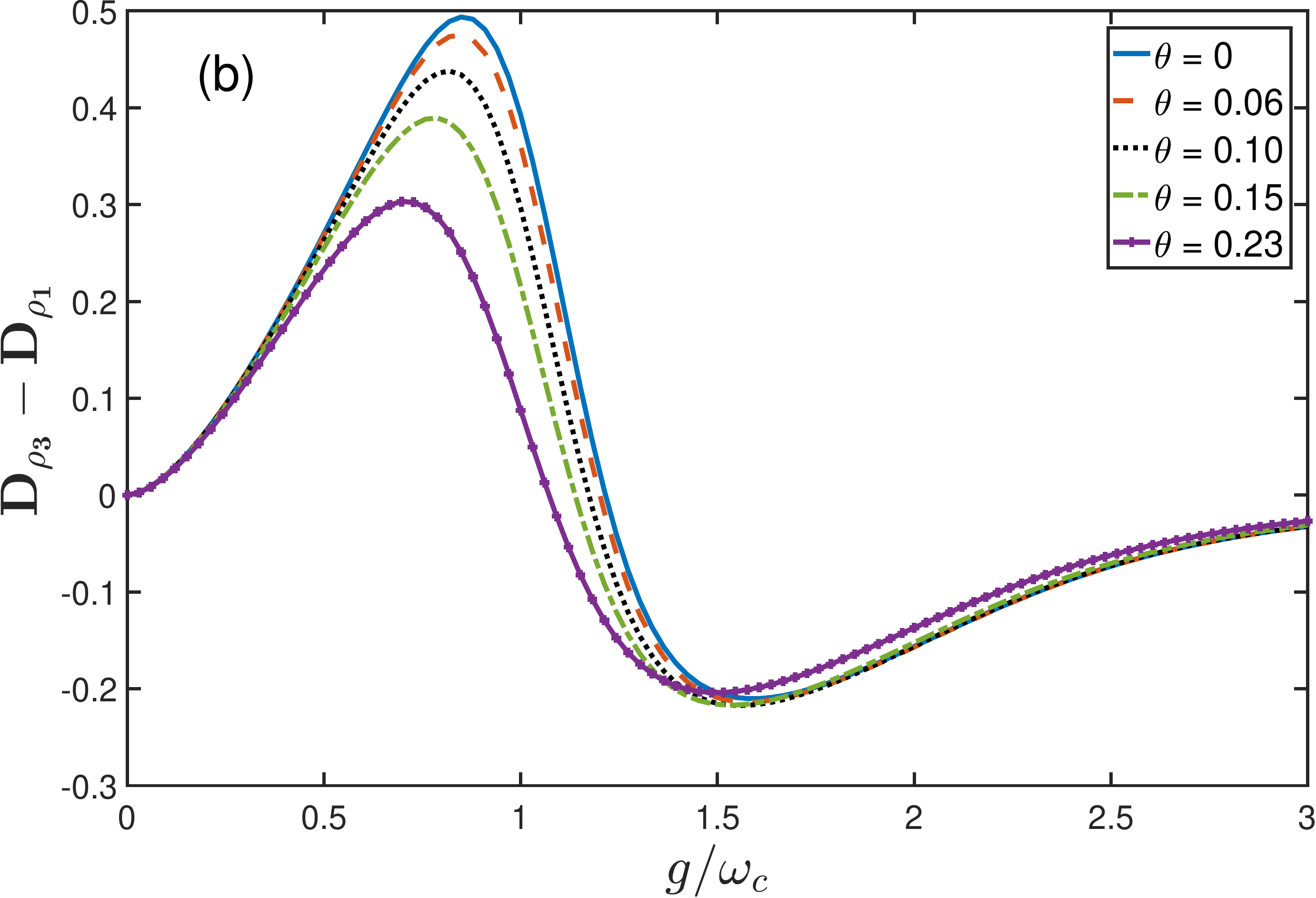}
	\end{subfigure}%
	\caption{ Difference of quantum correlations between states (a) $\rho_{4}$ and $\rho_{1}$ which are the initial and final states of the hot bath stage. And (b) states $\rho_{3}$ and $\rho_{1}$, which are the cold thermal state and hot thermal state respectively, as a function of coupling parameter $g/\omega_{c}$ for different values of the mixing angle $\theta$. We have used $\omega_{h} = 2\omega$ and $\omega_{c} = \omega$, $T_{c} = 19$ mK and $T_{h} = 9T_{c}$.}
	\label{fig:Corr}
\end{figure}
\section{Anharmonicity and engine operation}  \label{Anharmonicity}
It is known that for a working substance described by the quantum Rabi model ($\theta=0$) the QHE can experience a transition from a heat engine into a refrigerator depending on the light-matter coupling \cite{Hardal2015}. However it is not yet completely clear what properties of this working substance lead to such behaviour. In this section we show how the anharmonicity and degeneracy of the energy spectrum of the low energy levels of the quantum Rabi model together with the adiabatic process explain this positive to negative work transition. \par
  \begin{figure}[t!] 
	\centering
	\includegraphics[width=1\linewidth]{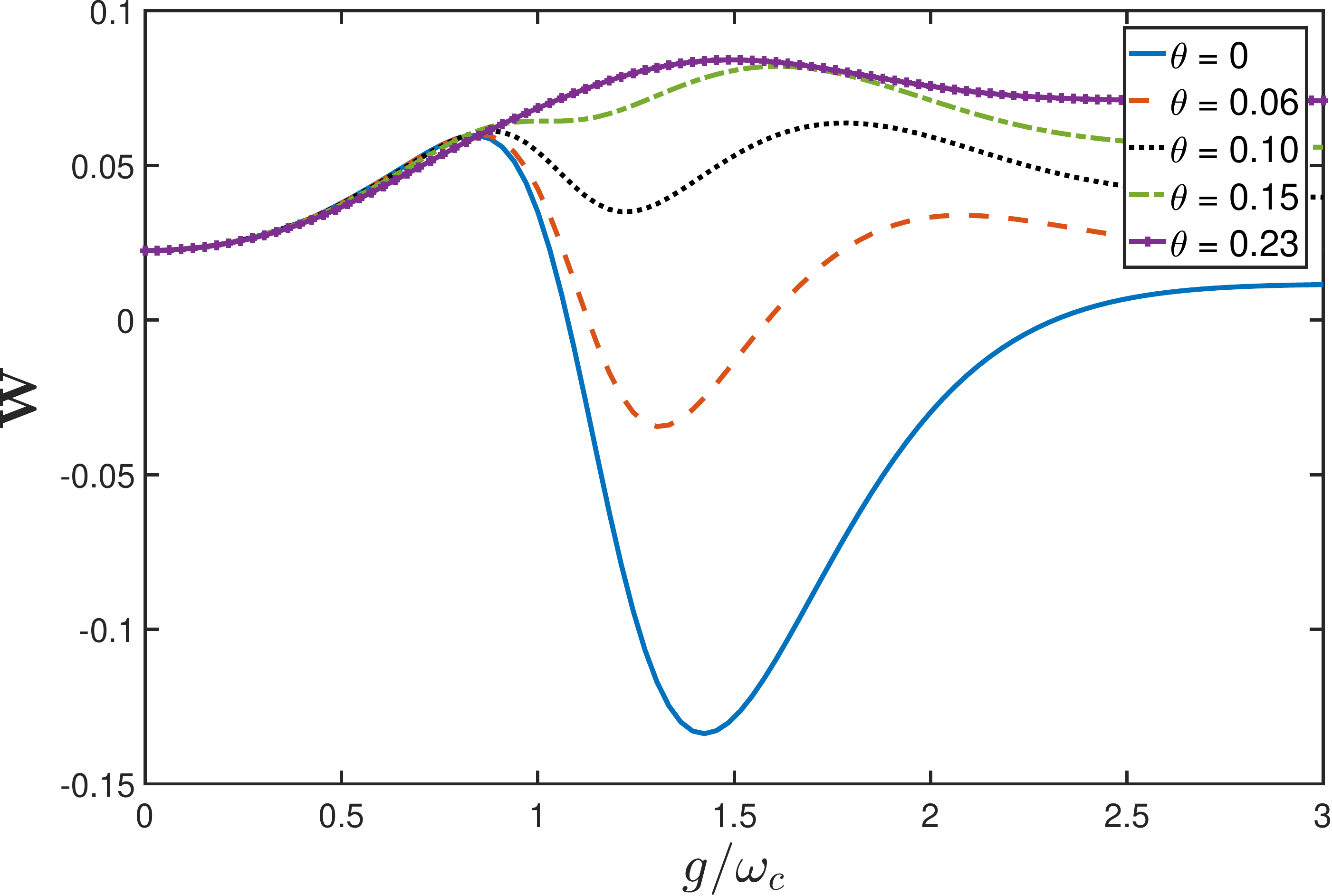}
	\caption{Work output of the engine versus the interaction parameter $g/\omega_{c}$ for different values of the mixing angle $\theta$. We have used $\omega_{h} = 2\omega$ and $\omega_{c} = \omega$, $T_{c} = 19$ mK and $T_{h} = 9T_{c}$.}
	\label{fig:work}
\end{figure}
 In Fig.~\ref{fig:work1} we show the total work output (black line) and the work contribution per energy level $W_{n}$ for the four lowest states. In this numerical calculation we chose $\omega_{h} = 2\omega_{c}$ such that the ratio $g/\omega$ of the hot Hamiltonian is half of the ratio of the cold Hamiltonian. The ground state has no contribution on the work extraction so we have $W_{0}=0$, this is because Eq.(\ref{Work}) depends only on the difference of energy levels  of $H_{h}$ and of $H_{c}$ relative to their own ground state. We can see from Fig.~\ref{fig:work1} that the main contribution to the total work comes from the first excited state through $W_{1}$, while the second and third excited states have a small and always positive contribution.  We see that the change in operation regime from heat engine into refrigerator at $g/\omega_{c} \lesssim 1.2$ is owed only to the contribution of the first excited state. In addition, at $g/\omega_{c} > 2.5$ the work contribution of the first excited state $W_{1}$ decreases to zero while the contributions of the second and third excited states become dominant and take the total work into the positive regime.\par 
  To clarify the last discussion in Fig.\ref{fig:3}(a) we show the energy of the first excited state in the hot and cold case, as well as the thermal energy for both reservoirs, where the ground state energy has been set to zero (see appendix \ref{ap0}, Fig. \ref{fig:Rabispectrum} for the energy spectrum of the quantum Rabi model for higher states). We can see that $E_{1}^{h} $ is always greater than $E_{1}^{c}$ which means that the energy difference factor in $W_{1}$ is always positive, hence the negative value of $W_{1}$ for $g/\omega_{c} \gtrsim 1.2$ originates from the population difference factor of $W_{1}$. In Fig.~\ref{fig:3}(b) we plot the thermal population of $|E^{h}_{1}\rangle$ (red line) and $|E_{1}^{c}\rangle$ (blue line). As we can see, in the region $g/\omega_{c} \lesssim 1.2$ the population of the hot excited state is greater than that of the cold excited state, and we have $P_{1}(T_{h}) - P_{1}(T_{c}) > 0$, but for greater values of $g/\omega_{c}$  we have $P_{1}(T_{h}) - P_{1}(T_{c}) < 0$ which corresponds to the refrigeration regime.\par
 To understand the behavior of the population difference explained above, we look at Fig.\ref{fig:3}(a) which shows that the ratio $E_{1}^{c,(h)}/k_{B}T_{c,(h)}$ decreases monotonously as $g/\omega_{c}$ increases. Near the point of intersection of $E_{1}^{c,(h)}$ and $k_{B}T_{c,(h)}$ the thermal population will increase rapidly. This is the role that the anharmonicity and degeneracy between the ground and first excited state play on the behavior of the thermal population and is necessary to have $P_{1}(T_{h}) - P_{1}(T_{c}) < 0 $. However, this is not enough to achieve a refrigeration regime, the missing condition is provided by the adiabatic process. As mentioned before, the adiabatic process makes the ratio $g/\omega$ of the hot Hamiltonian to be half the ratio of the cold Hamiltonian, as consequence, $E_{1}^{h}$ intersects $k_{B}T_{h}$ at a greater value of $g/\omega_{c}$ as compared with its cold counterpart, see Fig.\ref{fig:3}(a). This leads to the negative value of the population difference term in Eq.(\ref{Work2}). These two factors, namely, the anharmonicity of the energy spectrum and the specifics of the adiabatic process, give rise to the refrigeration regime. \par 
  By following a similar procedure as in Refs. \cite{Irish2007,Yu2012analytical} we can obtain an approximated expression for the energy levels, see appendix \ref{ap1}, which allow us to derive an approximated positive work condition for $W_{1}$ on the ratio $g/\omega_{c}$
\begin{equation}
\frac{g}{\omega} < \sqrt{ \frac{1}{2} \frac{R^{2}}{R^{2} - 1}  \textnormal{ln}\bigg(\frac{1}{R} \frac{T_{h}}{T_{c}}\bigg)},
\label{posw}
\end{equation}
where $R = \omega_{h}/\omega_{c} < T_{h}/T_{c}$. From this relation we can obtain an intuition about the effect of the adiabatic process on positive and negative work regime. For a fixed temperature ratio the right-hand side of Eq.(\ref{posw}) is a decreasing function of $R$ and this in turn indicates that an adiabatic process with large $R$ will have a small  interval of $g/\omega_{c}$ where the extractable work is positive.
   \begin{figure}[t!] 
	\centering
	\includegraphics[width=1\linewidth]{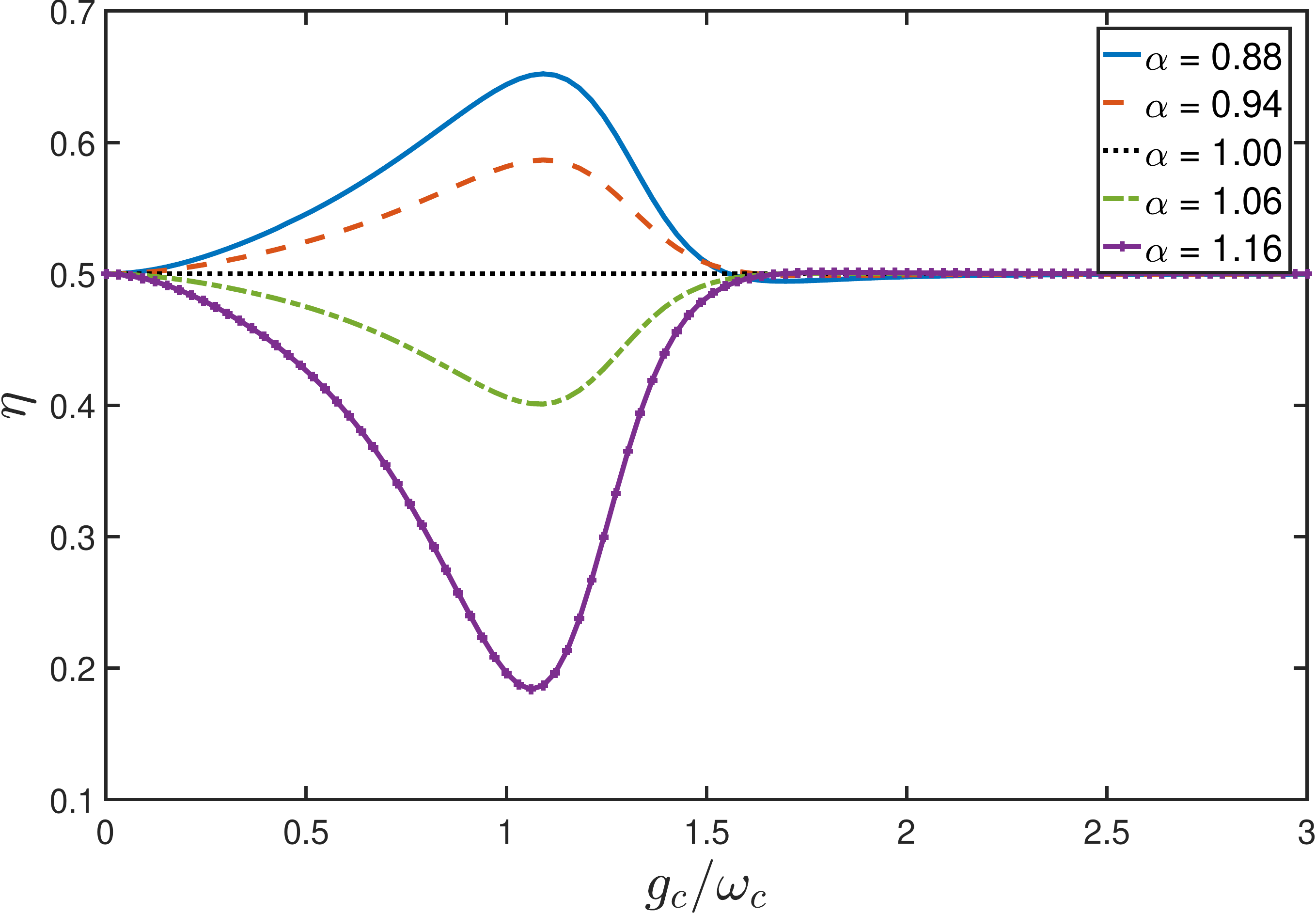}
	\caption{Efficiency of the QHE for different proportionality constant $\alpha$ versus the light-matter coupling $g_{c}/\omega_{c}$. We have used $\omega_{h} = 2\omega$ and $\omega_{c} = \omega$, $T_{c} = 19$ mK and $T_{h} = 9T_{c}$.}
	\label{fig:WEFF_CGW}
\end{figure}
  \section{Quantum correlations and work extraction}  \label{Qcorrelations}   
  We have learned that for the quantum Rabi model as working substance, the system can experience a transition in operation regime depending on the light-matter coupling strength. 
  An important issue to be considered  is the study  of correlations embedded in hybrid states of the quantum Rabi model, and the role they play in the operation regimes of the QHE.
  It has been suggested that the quantum correlations that are built at the end of the cold bath stage may be interpreted as a resource for enhancing work extraction \cite{Hardal2015}. The result that we present in what follows indicates that the difference of quantum correlations between the initial and final state of the hot bath stage is the quantity to be considered as a resource for harvested work.\par 
  We will quantify the quantum correlations between the two-level system and the field mode by means of quantum discord (QD). This measure allows to capture correlations that do not necessarily involve only quantum entanglement.\par 
\begin{figure}[t!] 
	\centering
	\includegraphics[width=1\linewidth]{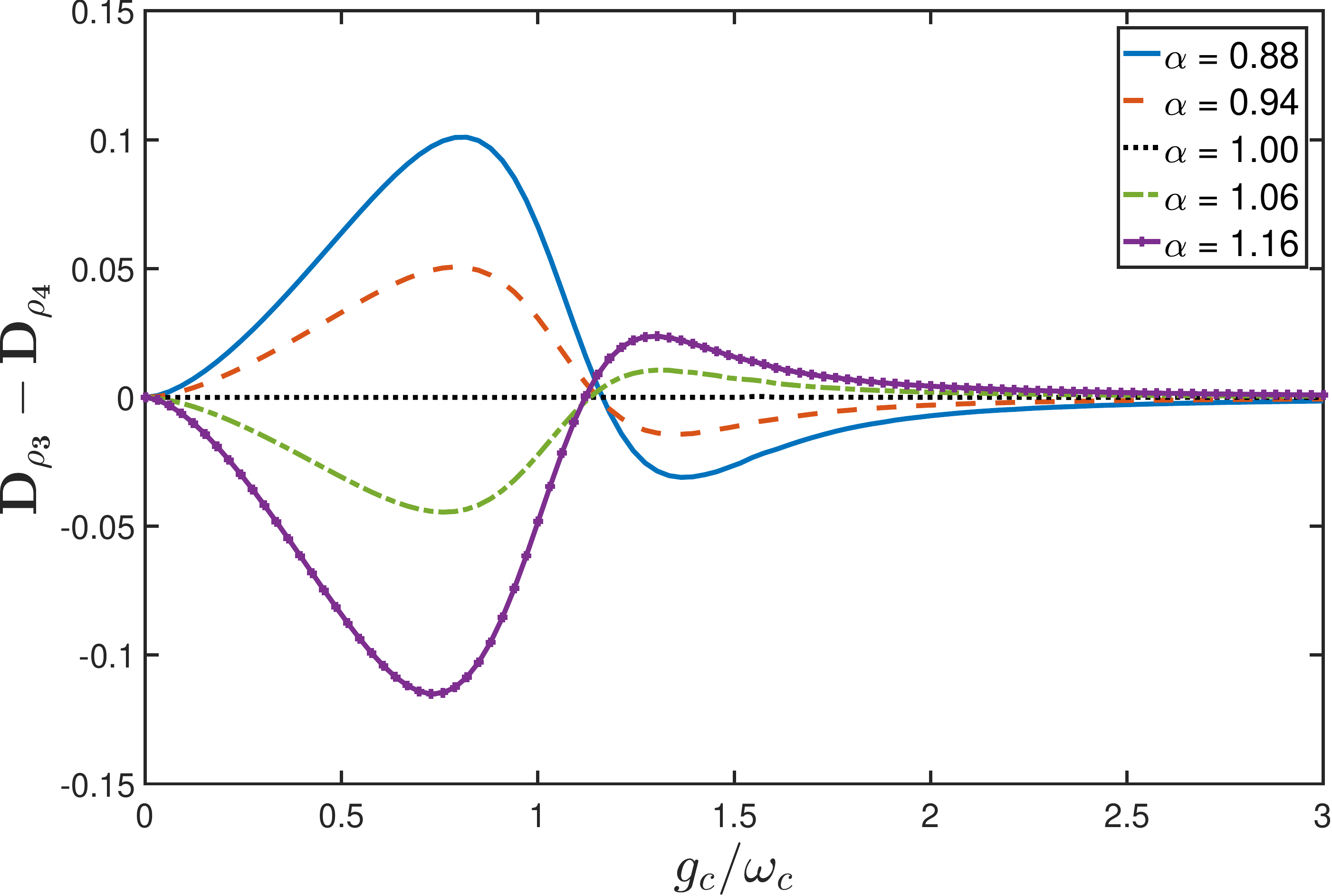}
	\caption{Difference of quantum correlations between states $\rho_{3}$ and $\rho_{4}$ for different proportionality constant $\alpha$, versus the light-matter coupling $g_{c}/\omega_{c}$. We have used $\omega_{h} = 2\omega$ and $\omega_{c} = \omega$, $T_{c} = 19$ mK and $T_{h} = 9T_{c}$.}
	\label{fig:QD_WEFF}
\end{figure} 
 The quantum discord in a bipartite $(A,B)$ system can be obtained as follows \cite{PhysRevA.79.042325}:\\
 \begin{equation}
 D_{A} = S(\rho_{A}) - S(\rho_{AB}) + \underset{\{\Pi_{j}^{A} \}}{\mathrm{min}} S(\rho_{B|\{\Pi_{j}^{A} \}}),
 \end{equation}
 where $S(\rho) = -\textnormal{tr}(\rho \textnormal{ln} \rho)$ is the von Neumann entropy, $\rho_{A,B}$ is the reduced density matrix for subsystem A or B, $\rho_{AB}$ is the density matrix of the complete system, and $\rho_{B|\{\Pi_{j}^{A} \}}$ is the state of the complete system after a projective measurement $\Pi_{j}^{A}$ is performed on subsystem $A$. In our case, the substystem A will be the  two-level system and subsystem B will be the field mode. Due to the size of the Hilbert space of the working substance we perform the projective measurements on the two-level system only. The strategy we use to calculate the optimal conditional entropy is the simulation technique used in Ref.\cite{Allende2015simulated} \par
 Which are the states whose quantum correlations should we consider as a resource for work extraction? We  know that in the hot bath stage quantum correlations are destroyed while in the cold bath stage correlations are created. On the other hand, the adiabatic compression and expansion processes can significantly change the amount of quantum correlations that are present in the thermal equilibrium states. We can intuitively think that during the hot bath stage the heat provided by the hot reservoir would have to be proportional to the amount of quantum correlations destroyed in this thermalization process. This suggests that the quantum correlations that we should consider are those present in the initial and final states of the hot bath stage. \par
In order to support the above claim, let us consider the generalized quantum Rabi model with $\theta \neq 0$. In Fig.~\ref{fig:Corr}(a) we plot the difference of quantum correlations between the initial and final states ($\rho_{4}$ and $\rho_{1}$) of the hot bath stage. In Fig.~\ref{fig:Corr}(b) we plot the difference of quantum correlations between the cold thermal state $\rho_{3}$ and the hot thermal state $\rho_{1}$. Also, in Fig.\ref{fig:work} we plot the extractable work $W$ for different values of the mixing angle $\theta$. If we compare Fig.~\ref{fig:Corr}(a) and Fig.\ref{fig:work} we can see that $D_{\rho_{4}} - D_{\rho_{1}}$ has a similar profile as the extractable work $W$ in the sense that maxima and minima can be equally identified. In this way, this difference of quantum correlations may be considered as a resource for work extraction. This is not the case for the difference of quantum correlations between the cold thermal state $\rho_{3}$ and the hot thermal state $\rho_{1}$ as shown in Fig.~\ref{fig:Corr}(b). Here, the  reduction in quantum correlations is no longer related to the behavior of the extractable work. We conclude that the quantum correlations which act as resource for work extraction are those of the initial and final state of the hot bath stage. \par  
\begin{figure}[t!] 
	\centering
	\includegraphics[width=1\linewidth]{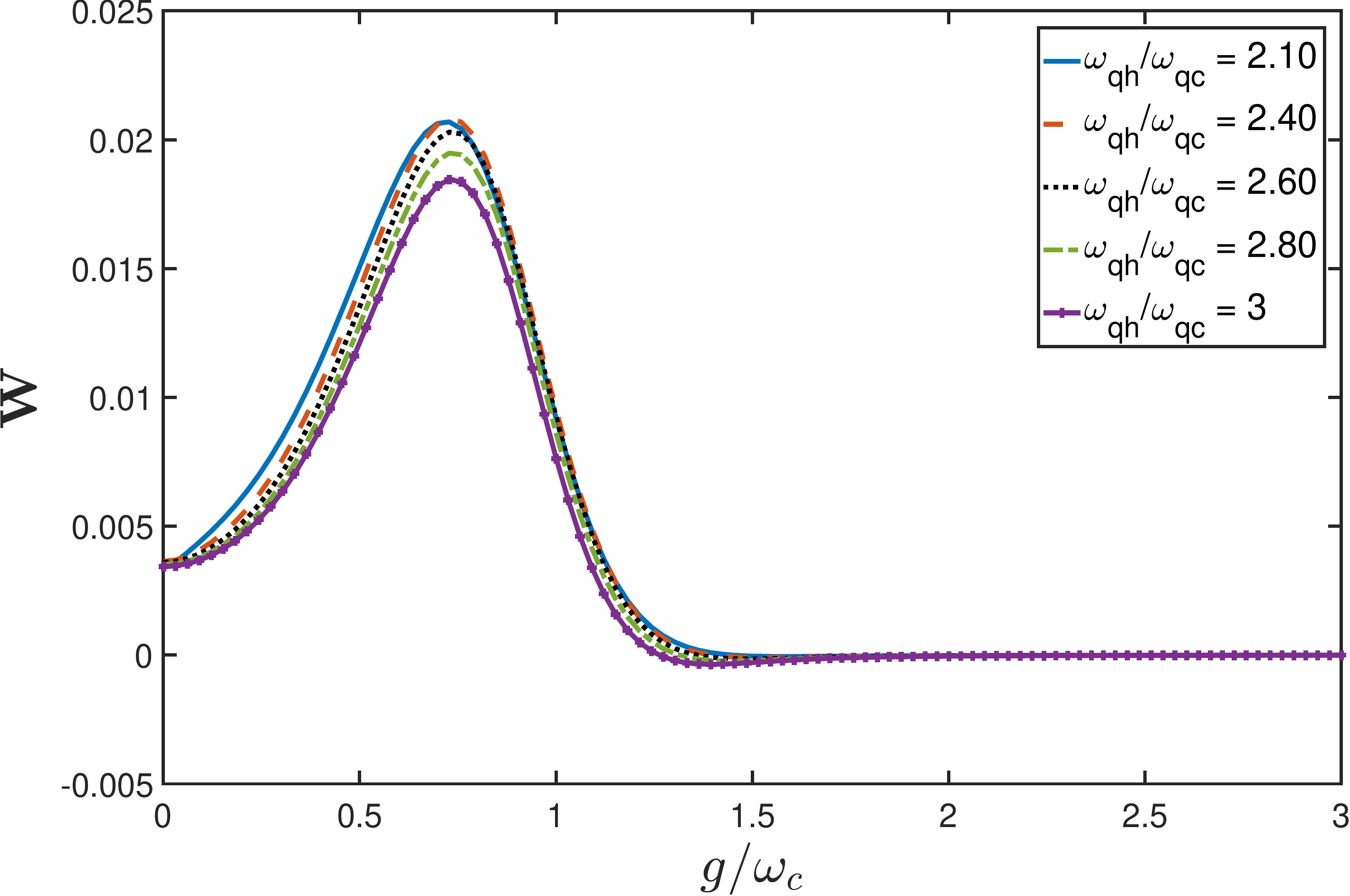}
	\caption{Work output of the engine versus the coupling parameter $g/\omega_{c}$ for different frequencies $\omega_{qh}$ of the hot qubit, where we have fixed $\omega_{qc}=0.5 \omega$. We have considered $T_{c} = 19$mK and $T_{h} = 4T_{c}$.}
	\label{fig:workCQ}
\end{figure}
 It is notheworthy that for the mixing angle $\theta= 0$, Fig.~\ref{fig:Corr}(a) and Fig.~\ref{fig:Corr}(b)  show the same behavior in accordance with the work extraction. In this case the compression stage does not change the quantum correlations significantly, so $\rho_{3}$ and $\rho_{4}$ have a very similar amount of quantum correlations. However, as the mixing angle increases  the quantum adiabatic processes change the amount of quantum correlations present in the cold thermal state  as evidenced by Fig.~\ref{fig:Corr}(a). This is the reason why the difference in quantum correlations $D_{\rho_{3}} - D_{\rho_{1}}$ cannot be considered as a resouce for work extraction.\par  
\section{Quantum Correlations and efficiency}  \label{Qcorrelations_eff}
Besides the characterization of the harvested work, a complete description of a QHE also requires the study of its efficiency.  As the work extraction arises from the adiabatic compression process, it is worthy to ask whether the change of quantum correlations induced by the compression stage is related to the efficiency of the QHE. In what follows, we show that the difference of quantum correlations in the compression stage can be considered a resource for enhanced efficiency of the QHE.\par 
We will consider an adiabatic process where we change simoultaneously the resonator frequency and the coupling strength parameter between the values $\{ \omega_{c},\omega_{h}\}$ and $\{ g_{c},g_{h}\}$ respectively. Specifically, we will consider an adiabatic process that sets $g_{h} = \alpha \frac{\omega_{h}}{\omega_{c}} g_{c}$, and study the difference of quantum correlations and efficiency of the QHE as a function of the coupling $g_{c}/\omega_{c}$ for different values of the parameter $\alpha$  and for a fixed $\omega_{h}/\omega_{c}$. Here we have included the change of the coupling parameter in our adiabatic process since it allows the hot Hamiltonian $H_{h}$ to be proportional to the cold Hamiltonian $H_{c}$ ($\alpha = 1$), and therefore, there is no change of quantum correlations in the compression stage.
 \begin{figure}[t!] 
	\centering
	\includegraphics[width=1\linewidth]{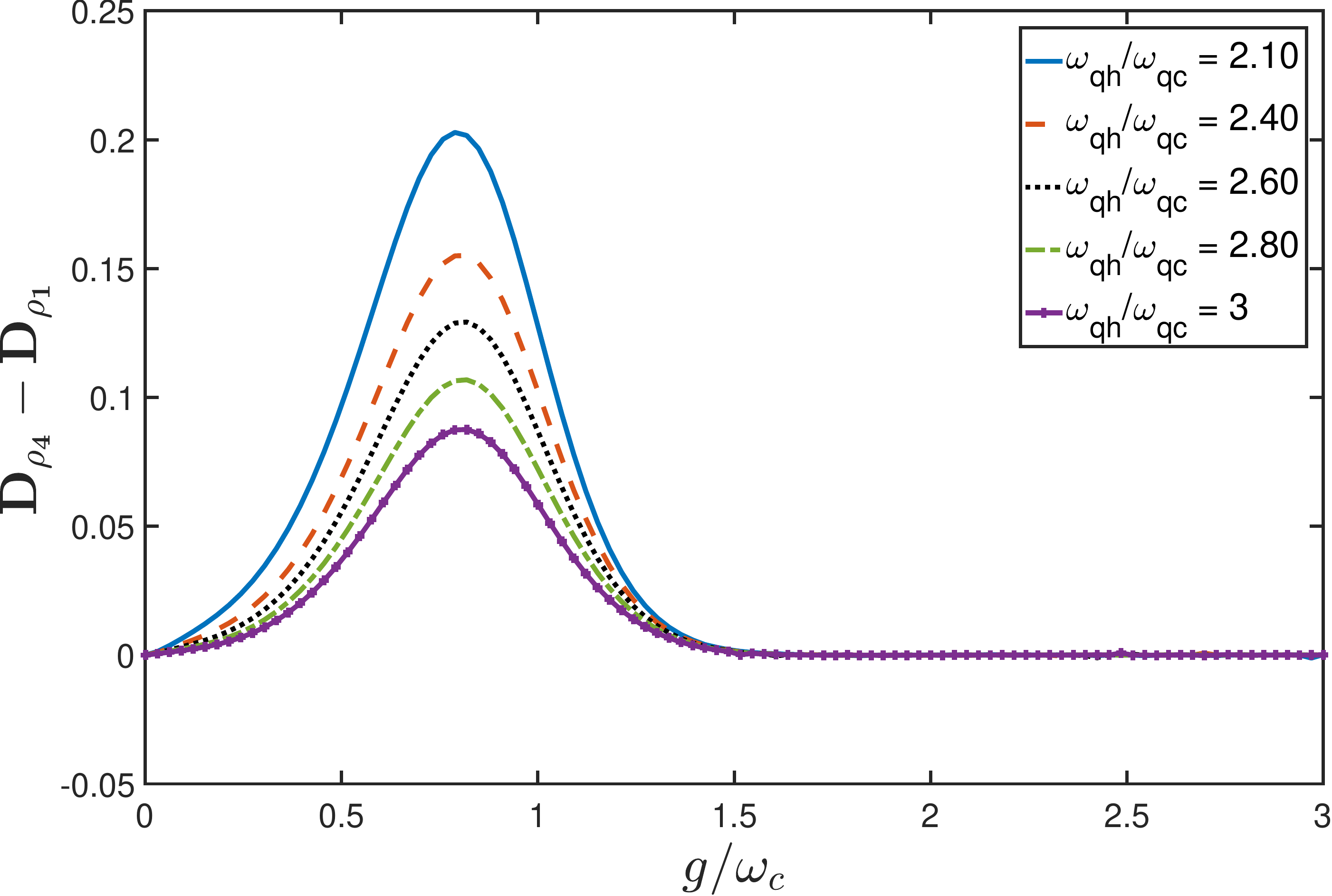}
	\caption{Difference of quantum correlations between the initial and final state of the hot bath stage as a function of coupling parameter $g/\omega_{c}$ for different frequencies $\omega_{qh}$ of the hot qubit, where we have fixed $\omega_{qc}=0.5 \omega$. We have used $\omega_{h} = \omega_{c} = \omega$ and $T_{c} = 19$mK and $T_{h} = 4T_{c}$.}
	\label{fig:EOFad}
\end{figure}
In Fig.\ref{fig:WEFF_CGW} we plot efficiency $\eta$ versus the coupling $g_{c}/\omega_{c}$ for different values of the proportionality constant $\alpha$, in this numerical calculation we use $\omega_{c} = \omega$, $\omega_{h} = 2 \omega$ and the mixing angle $\theta = 0$ for simplicity. Also in Fig.\ref{fig:QD_WEFF} we plot the difference of quantum correlations between the initial and final state of the compression stage. By comparing Fig.\ref{fig:WEFF_CGW} and Fig.\ref{fig:QD_WEFF}, we notice that when $\alpha = 1$, that is, when the hot Hamiltonian is proportional to the cold Hamiltonian, the light-matter coupling no longer affects the efficiency of the QHE, and there is no change of quantum correlations in the compression stage. For the case of $\alpha < 1$ we see that there is an optimal coupling region where the efficiency is maximized and it corresponds to the cases where there is a reduction in quantum correlations during the adiabatic compression state. On the other hand, when $\alpha > 1$ the coupling ratio $g_{c}/\omega_{c}$ has a negative impact on the efficiency in the region where the quantum correlation difference $D_{\rho_{3}} - D_{\rho_{4}}$ is negative. These observations allow us to conclude that the difference in quantum correlations between the initial and final states of the compression stage act as resource for enhanced engine efficiency. \par 
These results together with the previous section give us a broader picture of the role of quantum correlations in the performance of the QHE, the quantum correlation reduction during the hot isochoric stage acts as resource for harvested work while a quantum correlation reduction during the compression stage is an indicator of the behavior of the engine efficiency. 
\section{An optional adiabatic process}  \label{Optional}
 In the previous sections we have assumed the control of the coupling with respect to the cold resonator frequency  and we have described the regime of operation changing from heat engine into refrigerator. An alternative point of view to analyze our model is to consider, for example, to control only the qubit frequency which can be physically achieved in superconducting platforms by controlling the magnetic field threading a superconducting quantum interference device (SQUID) \cite{Paauw2009tuning,schwarz2013gradiometric}. In this case the adiabatic stages will only change the frequency of the qubit between the two values $\omega_{qc}$ and $\omega_{qh}$, while we fix the resonator frequency and coupling ratio $g/\omega$ for the Otto cycle.  For the sake of simplicity we will only consider the case of the quantum Rabi model, that is the mixing angle $\theta = 0$.\par 
 In Fig.\ref{fig:workCQ} we plot the work output of the engine for different frequencies $\omega_{qh}$ of the two-level system in the hot Hamiltonian, and in Fig. \ref{fig:EOFad} we plot the corresponding difference of quantum correlations measured as quantum discord between the initial and final state of the hot bath stage. As can be seen from both figures, we still see the same relation between extractable work and the reduction of quantum correlations that happens in the hot bath stage. In this case the quantum correlations in the hot thermal state are never greater than those present in $\rho_{3}$ and we observe only heat engine operation.\par 
 \begin{figure}[t!] 
	\centering
	\includegraphics[width=1\linewidth]{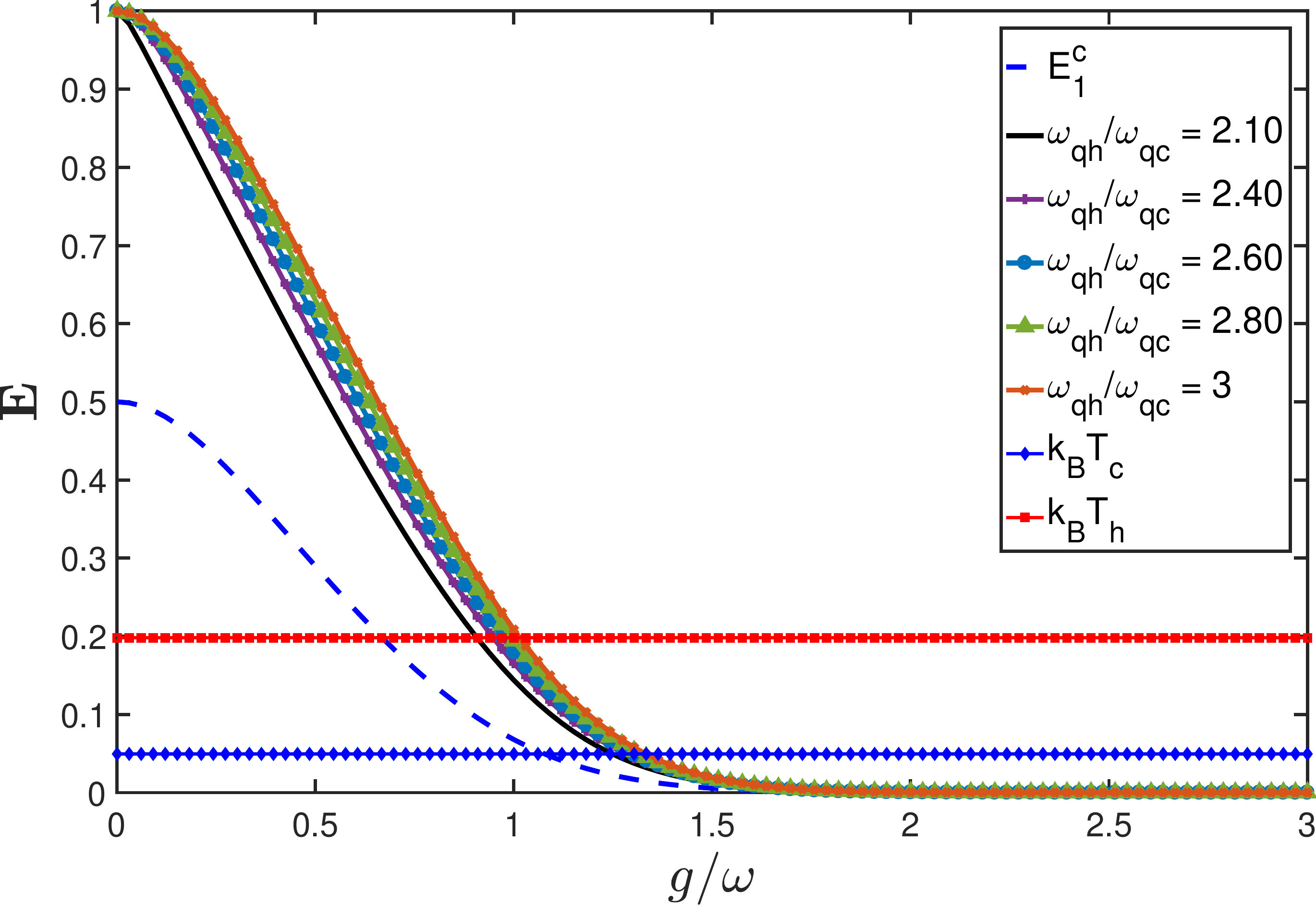}
	\caption{Energy of the first excited state $E_{1}$ for the cold Hamiltonian (blue dashed line), for the hot Hamiltonian for different values of the hot qubit frequency (solid lines) as a function of coupling parameter $g/\omega$. We also show the thermal energy of the hot reservoir (red squares) and cold reservoir (blue diamonds). Where we have used $\omega_{qc}=0.5 \omega$, $T_{c} = 19$mK and $T_{h} = 4T_{c}$.}
	\label{fig:Spectrum_qubit}
\end{figure} 
The absence of refrigeration regime, as compared with the results obtained in section \ref{Anharmonicity}, can be understood from Fig.\ref{fig:Spectrum_qubit} where we plot the energy of the first excited state of the quantum Rabi model for the different frequencies of the hot qubit. As can be seen from the figure, the energy of the hot first excited state intersects the hot thermal energy $k_{B}T_{h}$ in a smaller value of the ratio $g/\omega_{c}$ as compared with the cold first excited state and the cold thermal energy $k_{B}T_{c}$. As a consequence the thermal population of the hot first excited state will always be greater than its cold counterpart. Therefore, both terms of Eq.(\ref{Work2}) are positive and we can  conclude that this adiabatic process does not allow refrigeration regime.
\section{Conclusions} \label{conclusions}
We have shown how the anharmonicity and degeneracy of the spectrum of the quantum Rabi model give rise to an operation regime transition from heat engine into refrigerator as the light-matter coupling increases. By considering a generalized quantum Rabi model we found that  the quantum correlations reduction in the hot bath stage  and in the compression stage of the cycle can lead to enhanced positive work extraction and efficiency respectively. We have also shown that an alternative adiabatic process for the quantum Rabi model still maintains the relation we have shown between quantum correlations and positive work extraction, and may be implemented with the state-of-the-art circuit QED. Our results shed light on the search for optimal conditions in the performance of coupled quantum heat engines in relation to the properties of the energy spectrum and the capability of the working substance for embedding quantum correlations.\par 
\textit{Acknowledgements.} G.A.B acknowledges support from CONICYT Doctorado Nacional 21140587, F.A.-A acknowledges support CONICYT Doctorado Nacional 21140432, F.A.C.-L acknowledges support from  CEDENNA basal grant No. FB0807 and Direcci{\'o}n de Postgrado USACH, G.R acknowledges the support from FONDECYT under grant No. 1150653 and J. C. R acknowledges the support from FONDECYT under grant No. 1140194.
\bibliography{bib1}

\begin{thebibliography}{44}%
\makeatletter
\providecommand \@ifxundefined [1]{%
 \@ifx{#1\undefined}
}%
\providecommand \@ifnum [1]{%
 \ifnum #1\expandafter \@firstoftwo
 \else \expandafter \@secondoftwo
 \fi
}%
\providecommand \@ifx [1]{%
 \ifx #1\expandafter \@firstoftwo
 \else \expandafter \@secondoftwo
 \fi
}%
\providecommand \natexlab [1]{#1}%
\providecommand \enquote  [1]{``#1''}%
\providecommand \bibnamefont  [1]{#1}%
\providecommand \bibfnamefont [1]{#1}%
\providecommand \citenamefont [1]{#1}%
\providecommand \href@noop [0]{\@secondoftwo}%
\providecommand \href [0]{\begingroup \@sanitize@url \@href}%
\providecommand \@href[1]{\@@startlink{#1}\@@href}%
\providecommand \@@href[1]{\endgroup#1\@@endlink}%
\providecommand \@sanitize@url [0]{\catcode `\\12\catcode `\$12\catcode
  `\&12\catcode `\#12\catcode `\^12\catcode `\_12\catcode `\%12\relax}%
\providecommand \@@startlink[1]{}%
\providecommand \@@endlink[0]{}%
\providecommand \url  [0]{\begingroup\@sanitize@url \@url }%
\providecommand \@url [1]{\endgroup\@href {#1}{\urlprefix }}%
\providecommand \urlprefix  [0]{URL }%
\providecommand \Eprint [0]{\href }%
\providecommand \doibase [0]{http://dx.doi.org/}%
\providecommand \selectlanguage [0]{\@gobble}%
\providecommand \bibinfo  [0]{\@secondoftwo}%
\providecommand \bibfield  [0]{\@secondoftwo}%
\providecommand \translation [1]{[#1]}%
\providecommand \BibitemOpen [0]{}%
\providecommand \bibitemStop [0]{}%
\providecommand \bibitemNoStop [0]{.\EOS\space}%
\providecommand \EOS [0]{\spacefactor3000\relax}%
\providecommand \BibitemShut  [1]{\csname bibitem#1\endcsname}%
\let\auto@bib@innerbib\@empty
\bibitem [{\citenamefont {Scovil}\ and\ \citenamefont
  {Schulz-DuBois}(1959)}]{scovil1959three}%
  \BibitemOpen
  \bibfield  {author} {\bibinfo {author} {\bibfnamefont {H.~E.~D.}\
  \bibnamefont {Scovil}}\ and\ \bibinfo {author} {\bibfnamefont {E.~O.}\
  \bibnamefont {Schulz-DuBois}},\ }\href {\doibase 10.1103/PhysRevLett.2.262}
  {\bibfield  {journal} {\bibinfo  {journal} {Phys. Rev. Lett.}\ }\textbf
  {\bibinfo {volume} {2}},\ \bibinfo {pages} {262} (\bibinfo {year}
  {1959})}\BibitemShut {NoStop}%
\bibitem [{\citenamefont {Feldmann}\ \emph {et~al.}(1996)\citenamefont
  {Feldmann}, \citenamefont {Geva}, \citenamefont {Kosloff},\ and\
  \citenamefont {Salamon}}]{feldmann1996heat}%
  \BibitemOpen
  \bibfield  {author} {\bibinfo {author} {\bibfnamefont {T.}~\bibnamefont
  {Feldmann}}, \bibinfo {author} {\bibfnamefont {E.}~\bibnamefont {Geva}},
  \bibinfo {author} {\bibfnamefont {R.}~\bibnamefont {Kosloff}}, \ and\
  \bibinfo {author} {\bibfnamefont {P.}~\bibnamefont {Salamon}},\ }\href
  {\doibase 10.1119/1.18197} {\bibfield  {journal} {\bibinfo  {journal}
  {American Journal of Physics}\ }\textbf {\bibinfo {volume} {64}},\ \bibinfo
  {pages} {485} (\bibinfo {year} {1996})}\BibitemShut {NoStop}%
\bibitem [{\citenamefont {Feldmann}\ and\ \citenamefont
  {Kosloff}(2004)}]{feldmann2004characteristics}%
  \BibitemOpen
  \bibfield  {author} {\bibinfo {author} {\bibfnamefont {T.}~\bibnamefont
  {Feldmann}}\ and\ \bibinfo {author} {\bibfnamefont {R.}~\bibnamefont
  {Kosloff}},\ }\href {\doibase 10.1103/PhysRevE.70.046110} {\bibfield
  {journal} {\bibinfo  {journal} {Phys. Rev. E}\ }\textbf {\bibinfo {volume}
  {70}},\ \bibinfo {pages} {046110} (\bibinfo {year} {2004})}\BibitemShut
  {NoStop}%
\bibitem [{\citenamefont {Rezek}\ and\ \citenamefont
  {Kosloff}(2006)}]{rezek2006irreversible}%
  \BibitemOpen
  \bibfield  {author} {\bibinfo {author} {\bibfnamefont {Y.}~\bibnamefont
  {Rezek}}\ and\ \bibinfo {author} {\bibfnamefont {R.}~\bibnamefont
  {Kosloff}},\ }\href {http://stacks.iop.org/1367-2630/8/i=5/a=083} {\bibfield
  {journal} {\bibinfo  {journal} {New Journal of Physics}\ }\textbf {\bibinfo
  {volume} {8}},\ \bibinfo {pages} {83} (\bibinfo {year} {2006})}\BibitemShut
  {NoStop}%
\bibitem [{\citenamefont {Henrich}\ \emph {et~al.}(2007)\citenamefont
  {Henrich}, \citenamefont {Rempp},\ and\ \citenamefont
  {Mahler}}]{Henrich2007}%
  \BibitemOpen
  \bibfield  {author} {\bibinfo {author} {\bibfnamefont {M.~J.}\ \bibnamefont
  {Henrich}}, \bibinfo {author} {\bibfnamefont {F.}~\bibnamefont {Rempp}}, \
  and\ \bibinfo {author} {\bibfnamefont {G.}~\bibnamefont {Mahler}},\ }\href
  {\doibase 10.1140/epjst/e2007-00371-8} {\bibfield  {journal} {\bibinfo
  {journal} {The European Physical Journal Special Topics}\ }\textbf {\bibinfo
  {volume} {151}},\ \bibinfo {pages} {157} (\bibinfo {year}
  {2007})}\BibitemShut {NoStop}%
\bibitem [{\citenamefont {Quan}\ \emph {et~al.}(2007)\citenamefont {Quan},
  \citenamefont {Liu}, \citenamefont {Sun},\ and\ \citenamefont
  {Nori}}]{quan2007quantum}%
  \BibitemOpen
  \bibfield  {author} {\bibinfo {author} {\bibfnamefont {H.~T.}\ \bibnamefont
  {Quan}}, \bibinfo {author} {\bibfnamefont {Y.-x.}\ \bibnamefont {Liu}},
  \bibinfo {author} {\bibfnamefont {C.~P.}\ \bibnamefont {Sun}}, \ and\
  \bibinfo {author} {\bibfnamefont {F.}~\bibnamefont {Nori}},\ }\href {\doibase
  10.1103/PhysRevE.76.031105} {\bibfield  {journal} {\bibinfo  {journal} {Phys.
  Rev. E}\ }\textbf {\bibinfo {volume} {76}},\ \bibinfo {pages} {031105}
  (\bibinfo {year} {2007})}\BibitemShut {NoStop}%
\bibitem [{\citenamefont {He}\ \emph {et~al.}(2009)\citenamefont {He},
  \citenamefont {He},\ and\ \citenamefont {Tang}}]{he2009performance}%
  \BibitemOpen
  \bibfield  {author} {\bibinfo {author} {\bibfnamefont {J.}~\bibnamefont
  {He}}, \bibinfo {author} {\bibfnamefont {X.}~\bibnamefont {He}}, \ and\
  \bibinfo {author} {\bibfnamefont {W.}~\bibnamefont {Tang}},\ }\href {\doibase
  10.1007/s11433-009-0169-z} {\bibfield  {journal} {\bibinfo  {journal}
  {Science in China Series G: Physics, Mechanics and Astronomy}\ }\textbf
  {\bibinfo {volume} {52}},\ \bibinfo {pages} {1317} (\bibinfo {year}
  {2009})}\BibitemShut {NoStop}%
\bibitem [{\citenamefont {Thomas}\ \emph {et~al.}(2016)\citenamefont {Thomas},
  \citenamefont {Banik},\ and\ \citenamefont {Ghosh}}]{thomas2016performance}%
  \BibitemOpen
  \bibfield  {author} {\bibinfo {author} {\bibfnamefont {G.}~\bibnamefont
  {Thomas}}, \bibinfo {author} {\bibfnamefont {M.}~\bibnamefont {Banik}}, \
  and\ \bibinfo {author} {\bibfnamefont {S.}~\bibnamefont {Ghosh}},\
  }\href@noop {} {\bibfield  {journal} {\bibinfo  {journal} {arXiv preprint
  arXiv:1607.00994}\ } (\bibinfo {year} {2016})}\BibitemShut {NoStop}%
\bibitem [{\citenamefont {He}\ \emph {et~al.}(2002)\citenamefont {He},
  \citenamefont {Chen},\ and\ \citenamefont {Hua}}]{He2002}%
  \BibitemOpen
  \bibfield  {author} {\bibinfo {author} {\bibfnamefont {J.}~\bibnamefont
  {He}}, \bibinfo {author} {\bibfnamefont {J.}~\bibnamefont {Chen}}, \ and\
  \bibinfo {author} {\bibfnamefont {B.}~\bibnamefont {Hua}},\ }\href {\doibase
  10.1103/PhysRevE.65.036145} {\bibfield  {journal} {\bibinfo  {journal} {Phys.
  Rev. E}\ }\textbf {\bibinfo {volume} {65}},\ \bibinfo {pages} {036145}
  (\bibinfo {year} {2002})}\BibitemShut {NoStop}%
\bibitem [{\citenamefont {Geva}(1996)}]{Geva1996}%
  \BibitemOpen
  \bibfield  {author} {\bibinfo {author} {\bibfnamefont {E.}~\bibnamefont
  {Geva}},\ }\href {\doibase 10.1063/1.471453} {\bibfield  {journal} {\bibinfo
  {journal} {The Journal of Chemical Physics}\ }\textbf {\bibinfo {volume}
  {104}},\ \bibinfo {pages} {7681} (\bibinfo {year} {1996})}\BibitemShut
  {NoStop}%
\bibitem [{\citenamefont {Li}\ \emph {et~al.}(2007)\citenamefont {Li},
  \citenamefont {Wang}, \citenamefont {Sun},\ and\ \citenamefont
  {Yi}}]{li2007quantum}%
  \BibitemOpen
  \bibfield  {author} {\bibinfo {author} {\bibfnamefont {S.}~\bibnamefont
  {Li}}, \bibinfo {author} {\bibfnamefont {H.}~\bibnamefont {Wang}}, \bibinfo
  {author} {\bibfnamefont {Y.~D.}\ \bibnamefont {Sun}}, \ and\ \bibinfo
  {author} {\bibfnamefont {X.~X.}\ \bibnamefont {Yi}},\ }\href
  {http://stacks.iop.org/1751-8121/40/i=30/a=004} {\bibfield  {journal}
  {\bibinfo  {journal} {Journal of Physics A: Mathematical and Theoretical}\
  }\textbf {\bibinfo {volume} {40}},\ \bibinfo {pages} {8655} (\bibinfo {year}
  {2007})}\BibitemShut {NoStop}%
\bibitem [{\citenamefont {Bender}\ \emph {et~al.}(2002)\citenamefont {Bender},
  \citenamefont {Brody},\ and\ \citenamefont {Meister}}]{bender2002entropy}%
  \BibitemOpen
  \bibfield  {author} {\bibinfo {author} {\bibfnamefont {C.~M.}\ \bibnamefont
  {Bender}}, \bibinfo {author} {\bibfnamefont {D.~C.}\ \bibnamefont {Brody}}, \
  and\ \bibinfo {author} {\bibfnamefont {B.~K.}\ \bibnamefont {Meister}},\
  }\href {\doibase 10.1098/rspa.2001.0928} {\bibfield  {journal} {\bibinfo
  {journal} {Proceedings of the Royal Society of London A: Mathematical,
  Physical and Engineering Sciences}\ }\textbf {\bibinfo {volume} {458}},\
  \bibinfo {pages} {1519} (\bibinfo {year} {2002})}\BibitemShut {NoStop}%
\bibitem [{\citenamefont {Kosloff}\ and\ \citenamefont
  {Rezek}(2017)}]{Kosloff2017harmonic}%
  \BibitemOpen
  \bibfield  {author} {\bibinfo {author} {\bibfnamefont {R.}~\bibnamefont
  {Kosloff}}\ and\ \bibinfo {author} {\bibfnamefont {Y.}~\bibnamefont
  {Rezek}},\ }\href@noop {} {\bibfield  {journal} {\bibinfo  {journal}
  {Entropy}\ }\textbf {\bibinfo {volume} {19}} (\bibinfo {year}
  {2017})}\BibitemShut {NoStop}%
\bibitem [{\citenamefont {Altintas}\ \emph {et~al.}(2015)\citenamefont
  {Altintas}, \citenamefont {Hardal},\ and\ \citenamefont {M\"ustecapl\ifmmode
  \imath \else \i \fi{}o\ifmmode~\breve{g}\else \u{g}\fi{}lu}}]{Hardal2015}%
  \BibitemOpen
  \bibfield  {author} {\bibinfo {author} {\bibfnamefont {F.}~\bibnamefont
  {Altintas}}, \bibinfo {author} {\bibfnamefont {A.~U.~C.}\ \bibnamefont
  {Hardal}}, \ and\ \bibinfo {author} {\bibfnamefont {O.~E.}\ \bibnamefont
  {M\"ustecapl\ifmmode \imath \else \i \fi{}o\ifmmode~\breve{g}\else
  \u{g}\fi{}lu}},\ }\href {\doibase 10.1103/PhysRevA.91.023816} {\bibfield
  {journal} {\bibinfo  {journal} {Phys. Rev. A}\ }\textbf {\bibinfo {volume}
  {91}},\ \bibinfo {pages} {023816} (\bibinfo {year} {2015})}\BibitemShut
  {NoStop}%
\bibitem [{\citenamefont {Song}\ \emph {et~al.}(2016)\citenamefont {Song},
  \citenamefont {Singh}, \citenamefont {Zhang}, \citenamefont {Zhang},\ and\
  \citenamefont {Meystre}}]{song2016one}%
  \BibitemOpen
  \bibfield  {author} {\bibinfo {author} {\bibfnamefont {Q.}~\bibnamefont
  {Song}}, \bibinfo {author} {\bibfnamefont {S.}~\bibnamefont {Singh}},
  \bibinfo {author} {\bibfnamefont {K.}~\bibnamefont {Zhang}}, \bibinfo
  {author} {\bibfnamefont {W.}~\bibnamefont {Zhang}}, \ and\ \bibinfo {author}
  {\bibfnamefont {P.}~\bibnamefont {Meystre}},\ }\href {\doibase
  10.1103/PhysRevA.94.063852} {\bibfield  {journal} {\bibinfo  {journal} {Phys.
  Rev. A}\ }\textbf {\bibinfo {volume} {94}},\ \bibinfo {pages} {063852}
  (\bibinfo {year} {2016})}\BibitemShut {NoStop}%
\bibitem [{\citenamefont {Zhang}\ \emph {et~al.}(2007)\citenamefont {Zhang},
  \citenamefont {Liu}, \citenamefont {Chen},\ and\ \citenamefont
  {Li}}]{Zhang2007}%
  \BibitemOpen
  \bibfield  {author} {\bibinfo {author} {\bibfnamefont {T.}~\bibnamefont
  {Zhang}}, \bibinfo {author} {\bibfnamefont {W.-T.}\ \bibnamefont {Liu}},
  \bibinfo {author} {\bibfnamefont {P.-X.}\ \bibnamefont {Chen}}, \ and\
  \bibinfo {author} {\bibfnamefont {C.-Z.}\ \bibnamefont {Li}},\ }\href
  {\doibase 10.1103/PhysRevA.75.062102} {\bibfield  {journal} {\bibinfo
  {journal} {Phys. Rev. A}\ }\textbf {\bibinfo {volume} {75}},\ \bibinfo
  {pages} {062102} (\bibinfo {year} {2007})}\BibitemShut {NoStop}%
\bibitem [{\citenamefont {Zhang}(2008)}]{Zhang2008}%
  \BibitemOpen
  \bibfield  {author} {\bibinfo {author} {\bibfnamefont {G.~F.}\ \bibnamefont
  {Zhang}},\ }\href {\doibase 10.1140/epjd/e2008-00133-0} {\bibfield  {journal}
  {\bibinfo  {journal} {The European Physical Journal D}\ }\textbf {\bibinfo
  {volume} {49}},\ \bibinfo {pages} {123} (\bibinfo {year} {2008})}\BibitemShut
  {NoStop}%
\bibitem [{\citenamefont {Wang}\ \emph {et~al.}(2009)\citenamefont {Wang},
  \citenamefont {Liu},\ and\ \citenamefont {He}}]{Wang2009}%
  \BibitemOpen
  \bibfield  {author} {\bibinfo {author} {\bibfnamefont {H.}~\bibnamefont
  {Wang}}, \bibinfo {author} {\bibfnamefont {S.}~\bibnamefont {Liu}}, \ and\
  \bibinfo {author} {\bibfnamefont {J.}~\bibnamefont {He}},\ }\href {\doibase
  10.1103/PhysRevE.79.041113} {\bibfield  {journal} {\bibinfo  {journal} {Phys.
  Rev. E}\ }\textbf {\bibinfo {volume} {79}},\ \bibinfo {pages} {041113}
  (\bibinfo {year} {2009})}\BibitemShut {NoStop}%
\bibitem [{\citenamefont {Dillenschneider}\ and\ \citenamefont
  {Lutz}(2009)}]{dillenschneider2009}%
  \BibitemOpen
  \bibfield  {author} {\bibinfo {author} {\bibfnamefont {R.}~\bibnamefont
  {Dillenschneider}}\ and\ \bibinfo {author} {\bibfnamefont {E.}~\bibnamefont
  {Lutz}},\ }\href {http://stacks.iop.org/0295-5075/88/i=5/a=50003} {\bibfield
  {journal} {\bibinfo  {journal} {EPL (Europhysics Letters)}\ }\textbf
  {\bibinfo {volume} {88}},\ \bibinfo {pages} {50003} (\bibinfo {year}
  {2009})}\BibitemShut {NoStop}%
\bibitem [{\citenamefont {Altintas}\ \emph {et~al.}(2014)\citenamefont
  {Altintas}, \citenamefont {Hardal},\ and\ \citenamefont {M\"ustecapl\ifmmode
  \imath \else \i \fi{}o\ifmmode~\breve{g}\else \u{g}\fi{}lu}}]{Altintas2014}%
  \BibitemOpen
  \bibfield  {author} {\bibinfo {author} {\bibfnamefont {F.}~\bibnamefont
  {Altintas}}, \bibinfo {author} {\bibfnamefont {A.~U.~C.}\ \bibnamefont
  {Hardal}}, \ and\ \bibinfo {author} {\bibfnamefont {O.~E.}\ \bibnamefont
  {M\"ustecapl\ifmmode \imath \else \i \fi{}o\ifmmode~\breve{g}\else
  \u{g}\fi{}lu}},\ }\href {\doibase 10.1103/PhysRevE.90.032102} {\bibfield
  {journal} {\bibinfo  {journal} {Phys. Rev. E}\ }\textbf {\bibinfo {volume}
  {90}},\ \bibinfo {pages} {032102} (\bibinfo {year} {2014})}\BibitemShut
  {NoStop}%
\bibitem [{\citenamefont {Niemczyk}\ \emph {et~al.}(2010)\citenamefont
  {Niemczyk}, \citenamefont {Deppe}, \citenamefont {Huebl}, \citenamefont
  {Menzel}, \citenamefont {Hocke}, \citenamefont {Schwarz}, \citenamefont
  {Garcia-Ripoll}, \citenamefont {Zueco}, \citenamefont {H{\"u}mmer},
  \citenamefont {Solano} \emph {et~al.}}]{niemczyk2010circuit}%
  \BibitemOpen
  \bibfield  {author} {\bibinfo {author} {\bibfnamefont {T.}~\bibnamefont
  {Niemczyk}}, \bibinfo {author} {\bibfnamefont {F.}~\bibnamefont {Deppe}},
  \bibinfo {author} {\bibfnamefont {H.}~\bibnamefont {Huebl}}, \bibinfo
  {author} {\bibfnamefont {E.}~\bibnamefont {Menzel}}, \bibinfo {author}
  {\bibfnamefont {F.}~\bibnamefont {Hocke}}, \bibinfo {author} {\bibfnamefont
  {M.}~\bibnamefont {Schwarz}}, \bibinfo {author} {\bibfnamefont
  {J.}~\bibnamefont {Garcia-Ripoll}}, \bibinfo {author} {\bibfnamefont
  {D.}~\bibnamefont {Zueco}}, \bibinfo {author} {\bibfnamefont
  {T.}~\bibnamefont {H{\"u}mmer}}, \bibinfo {author} {\bibfnamefont
  {E.}~\bibnamefont {Solano}},  \emph {et~al.},\ }\href {\doibase
  10.1038/nphys1730} {\bibfield  {journal} {\bibinfo  {journal} {Nature
  Physics}\ }\textbf {\bibinfo {volume} {6}},\ \bibinfo {pages} {772} (\bibinfo
  {year} {2010})}\BibitemShut {NoStop}%
\bibitem [{\citenamefont {Forn-D\'{\i}az}\ \emph {et~al.}(2010)\citenamefont
  {Forn-D\'{\i}az}, \citenamefont {Lisenfeld}, \citenamefont {Marcos},
  \citenamefont {Garc\'{\i}a-Ripoll}, \citenamefont {Solano}, \citenamefont
  {Harmans},\ and\ \citenamefont {Mooij}}]{FornDiaz2010BSS}%
  \BibitemOpen
  \bibfield  {author} {\bibinfo {author} {\bibfnamefont {P.}~\bibnamefont
  {Forn-D\'{\i}az}}, \bibinfo {author} {\bibfnamefont {J.}~\bibnamefont
  {Lisenfeld}}, \bibinfo {author} {\bibfnamefont {D.}~\bibnamefont {Marcos}},
  \bibinfo {author} {\bibfnamefont {J.~J.}\ \bibnamefont {Garc\'{\i}a-Ripoll}},
  \bibinfo {author} {\bibfnamefont {E.}~\bibnamefont {Solano}}, \bibinfo
  {author} {\bibfnamefont {C.~J. P.~M.}\ \bibnamefont {Harmans}}, \ and\
  \bibinfo {author} {\bibfnamefont {J.~E.}\ \bibnamefont {Mooij}},\ }\href
  {\doibase 10.1103/PhysRevLett.105.237001} {\bibfield  {journal} {\bibinfo
  {journal} {Phys. Rev. Lett.}\ }\textbf {\bibinfo {volume} {105}},\ \bibinfo
  {pages} {237001} (\bibinfo {year} {2010})}\BibitemShut {NoStop}%
\bibitem [{\citenamefont {Bourassa}\ \emph {et~al.}(2009)\citenamefont
  {Bourassa}, \citenamefont {Gambetta}, \citenamefont {Abdumalikov},
  \citenamefont {Astafiev}, \citenamefont {Nakamura},\ and\ \citenamefont
  {Blais}}]{Bourassa2009USC}%
  \BibitemOpen
  \bibfield  {author} {\bibinfo {author} {\bibfnamefont {J.}~\bibnamefont
  {Bourassa}}, \bibinfo {author} {\bibfnamefont {J.~M.}\ \bibnamefont
  {Gambetta}}, \bibinfo {author} {\bibfnamefont {A.~A.}\ \bibnamefont
  {Abdumalikov}}, \bibinfo {author} {\bibfnamefont {O.}~\bibnamefont
  {Astafiev}}, \bibinfo {author} {\bibfnamefont {Y.}~\bibnamefont {Nakamura}},
  \ and\ \bibinfo {author} {\bibfnamefont {A.}~\bibnamefont {Blais}},\ }\href
  {\doibase 10.1103/PhysRevA.80.032109} {\bibfield  {journal} {\bibinfo
  {journal} {Phys. Rev. A}\ }\textbf {\bibinfo {volume} {80}},\ \bibinfo
  {pages} {032109} (\bibinfo {year} {2009})}\BibitemShut {NoStop}%
\bibitem [{\citenamefont {Yoshihara}\ \emph
  {et~al.}(2017{\natexlab{a}})\citenamefont {Yoshihara}, \citenamefont {Fuse},
  \citenamefont {Ashhab}, \citenamefont {Kakuyanagi}, \citenamefont {Saito},\
  and\ \citenamefont {Semba}}]{yoshihara2017DSC}%
  \BibitemOpen
  \bibfield  {author} {\bibinfo {author} {\bibfnamefont {F.}~\bibnamefont
  {Yoshihara}}, \bibinfo {author} {\bibfnamefont {T.}~\bibnamefont {Fuse}},
  \bibinfo {author} {\bibfnamefont {S.}~\bibnamefont {Ashhab}}, \bibinfo
  {author} {\bibfnamefont {K.}~\bibnamefont {Kakuyanagi}}, \bibinfo {author}
  {\bibfnamefont {S.}~\bibnamefont {Saito}}, \ and\ \bibinfo {author}
  {\bibfnamefont {K.}~\bibnamefont {Semba}},\ }\href {\doibase
  10.1038/nphys3906} {\bibfield  {journal} {\bibinfo  {journal} {Nature
  Physics}\ }\textbf {\bibinfo {volume} {13}},\ \bibinfo {pages} {44} (\bibinfo
  {year} {2017}{\natexlab{a}})}\BibitemShut {NoStop}%
\bibitem [{\citenamefont {Forn-D{\'\i}az}\ \emph {et~al.}(2017)\citenamefont
  {Forn-D{\'\i}az}, \citenamefont {Garc{\'\i}a-Ripoll}, \citenamefont
  {Peropadre}, \citenamefont {Orgiazzi}, \citenamefont {Yurtalan},
  \citenamefont {Belyansky}, \citenamefont {Wilson},\ and\ \citenamefont
  {Lupascu}}]{forn2017ultrastrong}%
  \BibitemOpen
  \bibfield  {author} {\bibinfo {author} {\bibfnamefont {P.}~\bibnamefont
  {Forn-D{\'\i}az}}, \bibinfo {author} {\bibfnamefont {J.}~\bibnamefont
  {Garc{\'\i}a-Ripoll}}, \bibinfo {author} {\bibfnamefont {B.}~\bibnamefont
  {Peropadre}}, \bibinfo {author} {\bibfnamefont {J.-L.}\ \bibnamefont
  {Orgiazzi}}, \bibinfo {author} {\bibfnamefont {M.}~\bibnamefont {Yurtalan}},
  \bibinfo {author} {\bibfnamefont {R.}~\bibnamefont {Belyansky}}, \bibinfo
  {author} {\bibfnamefont {C.}~\bibnamefont {Wilson}}, \ and\ \bibinfo {author}
  {\bibfnamefont {A.}~\bibnamefont {Lupascu}},\ }\href {\doibase
  10.1038/nphys3905} {\bibfield  {journal} {\bibinfo  {journal} {Nature
  Physics}\ }\textbf {\bibinfo {volume} {13}},\ \bibinfo {pages} {39} (\bibinfo
  {year} {2017})}\BibitemShut {NoStop}%
\bibitem [{\citenamefont {Yoshihara}\ \emph
  {et~al.}(2017{\natexlab{b}})\citenamefont {Yoshihara}, \citenamefont {Fuse},
  \citenamefont {Ashhab}, \citenamefont {Kakuyanagi}, \citenamefont {Saito},\
  and\ \citenamefont {Semba}}]{Yoshihara2017characteristic}%
  \BibitemOpen
  \bibfield  {author} {\bibinfo {author} {\bibfnamefont {F.}~\bibnamefont
  {Yoshihara}}, \bibinfo {author} {\bibfnamefont {T.}~\bibnamefont {Fuse}},
  \bibinfo {author} {\bibfnamefont {S.}~\bibnamefont {Ashhab}}, \bibinfo
  {author} {\bibfnamefont {K.}~\bibnamefont {Kakuyanagi}}, \bibinfo {author}
  {\bibfnamefont {S.}~\bibnamefont {Saito}}, \ and\ \bibinfo {author}
  {\bibfnamefont {K.}~\bibnamefont {Semba}},\ }\href {\doibase
  10.1103/PhysRevA.95.053824} {\bibfield  {journal} {\bibinfo  {journal} {Phys.
  Rev. A}\ }\textbf {\bibinfo {volume} {95}},\ \bibinfo {pages} {053824}
  (\bibinfo {year} {2017}{\natexlab{b}})}\BibitemShut {NoStop}%
\bibitem [{\citenamefont {Rabi}(1937)}]{Rabi1937}%
  \BibitemOpen
  \bibfield  {author} {\bibinfo {author} {\bibfnamefont {I.~I.}\ \bibnamefont
  {Rabi}},\ }\href {\doibase 10.1103/PhysRev.51.652} {\bibfield  {journal}
  {\bibinfo  {journal} {Phys. Rev.}\ }\textbf {\bibinfo {volume} {51}},\
  \bibinfo {pages} {652} (\bibinfo {year} {1937})}\BibitemShut {NoStop}%
\bibitem [{\citenamefont {Braak}(2011)}]{Braak2011}%
  \BibitemOpen
  \bibfield  {author} {\bibinfo {author} {\bibfnamefont {D.}~\bibnamefont
  {Braak}},\ }\href {\doibase 10.1103/PhysRevLett.107.100401} {\bibfield
  {journal} {\bibinfo  {journal} {Phys. Rev. Lett.}\ }\textbf {\bibinfo
  {volume} {107}},\ \bibinfo {pages} {100401} (\bibinfo {year}
  {2011})}\BibitemShut {NoStop}%
\bibitem [{\citenamefont {Nataf}\ and\ \citenamefont
  {Ciuti}(2011)}]{Nataf2011}%
  \BibitemOpen
  \bibfield  {author} {\bibinfo {author} {\bibfnamefont {P.}~\bibnamefont
  {Nataf}}\ and\ \bibinfo {author} {\bibfnamefont {C.}~\bibnamefont {Ciuti}},\
  }\href {\doibase 10.1103/PhysRevLett.107.190402} {\bibfield  {journal}
  {\bibinfo  {journal} {Phys. Rev. Lett.}\ }\textbf {\bibinfo {volume} {107}},\
  \bibinfo {pages} {190402} (\bibinfo {year} {2011})}\BibitemShut {NoStop}%
\bibitem [{\citenamefont {Romero}\ \emph {et~al.}(2012)\citenamefont {Romero},
  \citenamefont {Ballester}, \citenamefont {Wang}, \citenamefont {Scarani},\
  and\ \citenamefont {Solano}}]{Romero2012ultrafast}%
  \BibitemOpen
  \bibfield  {author} {\bibinfo {author} {\bibfnamefont {G.}~\bibnamefont
  {Romero}}, \bibinfo {author} {\bibfnamefont {D.}~\bibnamefont {Ballester}},
  \bibinfo {author} {\bibfnamefont {Y.~M.}\ \bibnamefont {Wang}}, \bibinfo
  {author} {\bibfnamefont {V.}~\bibnamefont {Scarani}}, \ and\ \bibinfo
  {author} {\bibfnamefont {E.}~\bibnamefont {Solano}},\ }\href {\doibase
  10.1103/PhysRevLett.108.120501} {\bibfield  {journal} {\bibinfo  {journal}
  {Phys. Rev. Lett.}\ }\textbf {\bibinfo {volume} {108}},\ \bibinfo {pages}
  {120501} (\bibinfo {year} {2012})}\BibitemShut {NoStop}%
\bibitem [{\citenamefont {Kyaw}\ \emph
  {et~al.}(2015{\natexlab{a}})\citenamefont {Kyaw}, \citenamefont {Felicetti},
  \citenamefont {Romero}, \citenamefont {Solano},\ and\ \citenamefont
  {Kwek}}]{kyaw2015scalable}%
  \BibitemOpen
  \bibfield  {author} {\bibinfo {author} {\bibfnamefont {T.~H.}\ \bibnamefont
  {Kyaw}}, \bibinfo {author} {\bibfnamefont {S.}~\bibnamefont {Felicetti}},
  \bibinfo {author} {\bibfnamefont {G.}~\bibnamefont {Romero}}, \bibinfo
  {author} {\bibfnamefont {E.}~\bibnamefont {Solano}}, \ and\ \bibinfo {author}
  {\bibfnamefont {L.-C.}\ \bibnamefont {Kwek}},\ }\href {\doibase
  10.1038/srep08621} {\bibfield  {journal} {\bibinfo  {journal} {Scientific
  Reports}\ }\textbf {\bibinfo {volume} {5}},\ \bibinfo {pages} {8621}
  (\bibinfo {year} {2015}{\natexlab{a}})}\BibitemShut {NoStop}%
\bibitem [{\citenamefont {Kyaw}\ \emph
  {et~al.}(2015{\natexlab{b}})\citenamefont {Kyaw}, \citenamefont
  {Herrera-Mart\'{\i}}, \citenamefont {Solano}, \citenamefont {Romero},\ and\
  \citenamefont {Kwek}}]{Kyaw2015QEC}%
  \BibitemOpen
  \bibfield  {author} {\bibinfo {author} {\bibfnamefont {T.~H.}\ \bibnamefont
  {Kyaw}}, \bibinfo {author} {\bibfnamefont {D.~A.}\ \bibnamefont
  {Herrera-Mart\'{\i}}}, \bibinfo {author} {\bibfnamefont {E.}~\bibnamefont
  {Solano}}, \bibinfo {author} {\bibfnamefont {G.}~\bibnamefont {Romero}}, \
  and\ \bibinfo {author} {\bibfnamefont {L.-C.}\ \bibnamefont {Kwek}},\ }\href
  {\doibase 10.1103/PhysRevB.91.064503} {\bibfield  {journal} {\bibinfo
  {journal} {Phys. Rev. B}\ }\textbf {\bibinfo {volume} {91}},\ \bibinfo
  {pages} {064503} (\bibinfo {year} {2015}{\natexlab{b}})}\BibitemShut
  {NoStop}%
\bibitem [{\citenamefont {Joshi}\ \emph {et~al.}(2017)\citenamefont {Joshi},
  \citenamefont {Irish},\ and\ \citenamefont {Spiller}}]{joshi2017qubit}%
  \BibitemOpen
  \bibfield  {author} {\bibinfo {author} {\bibfnamefont {C.}~\bibnamefont
  {Joshi}}, \bibinfo {author} {\bibfnamefont {E.~K.}\ \bibnamefont {Irish}}, \
  and\ \bibinfo {author} {\bibfnamefont {T.~P.}\ \bibnamefont {Spiller}},\
  }\href {\doibase 10.1038/srep45587} {\bibfield  {journal} {\bibinfo
  {journal} {Scientific Reports}\ }\textbf {\bibinfo {volume} {7}},\ \bibinfo
  {pages} {45587} (\bibinfo {year} {2017})}\BibitemShut {NoStop}%
\bibitem [{\citenamefont {Wolf}\ \emph {et~al.}(2013)\citenamefont {Wolf},
  \citenamefont {Vallone}, \citenamefont {Romero}, \citenamefont {Kollar},
  \citenamefont {Solano},\ and\ \citenamefont {Braak}}]{Wolf2013}%
  \BibitemOpen
  \bibfield  {author} {\bibinfo {author} {\bibfnamefont {F.~A.}\ \bibnamefont
  {Wolf}}, \bibinfo {author} {\bibfnamefont {F.}~\bibnamefont {Vallone}},
  \bibinfo {author} {\bibfnamefont {G.}~\bibnamefont {Romero}}, \bibinfo
  {author} {\bibfnamefont {M.}~\bibnamefont {Kollar}}, \bibinfo {author}
  {\bibfnamefont {E.}~\bibnamefont {Solano}}, \ and\ \bibinfo {author}
  {\bibfnamefont {D.}~\bibnamefont {Braak}},\ }\href {\doibase
  10.1103/PhysRevA.87.023835} {\bibfield  {journal} {\bibinfo  {journal} {Phys.
  Rev. A}\ }\textbf {\bibinfo {volume} {87}},\ \bibinfo {pages} {023835}
  (\bibinfo {year} {2013})}\BibitemShut {NoStop}%
\bibitem [{\citenamefont {Rossatto}\ \emph {et~al.}(2016)\citenamefont
  {Rossatto}, \citenamefont {Villas-B{\^o}as}, \citenamefont {Sanz},\ and\
  \citenamefont {Solano}}]{Rossatto2016}%
  \BibitemOpen
  \bibfield  {author} {\bibinfo {author} {\bibfnamefont {D.~Z.}\ \bibnamefont
  {Rossatto}}, \bibinfo {author} {\bibfnamefont {C.~J.}\ \bibnamefont
  {Villas-B{\^o}as}}, \bibinfo {author} {\bibfnamefont {M.}~\bibnamefont
  {Sanz}}, \ and\ \bibinfo {author} {\bibfnamefont {E.}~\bibnamefont
  {Solano}},\ }\href@noop {} {\bibfield  {journal} {\bibinfo  {journal} {arXiv
  preprint arXiv:1612.03090}\ } (\bibinfo {year} {2016})}\BibitemShut {NoStop}%
\bibitem [{\citenamefont {Jaynes}\ and\ \citenamefont
  {Cummings}(1963)}]{JC1963}%
  \BibitemOpen
  \bibfield  {author} {\bibinfo {author} {\bibfnamefont {E.~T.}\ \bibnamefont
  {Jaynes}}\ and\ \bibinfo {author} {\bibfnamefont {F.~W.}\ \bibnamefont
  {Cummings}},\ }\href {\doibase 10.1109/PROC.1963.1664} {\bibfield  {journal}
  {\bibinfo  {journal} {Proceedings of the IEEE}\ }\textbf {\bibinfo {volume}
  {51}},\ \bibinfo {pages} {89} (\bibinfo {year} {1963})}\BibitemShut {NoStop}%
\bibitem [{\citenamefont {Casanova}\ \emph {et~al.}(2010)\citenamefont
  {Casanova}, \citenamefont {Romero}, \citenamefont {Lizuain}, \citenamefont
  {Garc\'{\i}a-Ripoll},\ and\ \citenamefont {Solano}}]{Casanova2010DSC}%
  \BibitemOpen
  \bibfield  {author} {\bibinfo {author} {\bibfnamefont {J.}~\bibnamefont
  {Casanova}}, \bibinfo {author} {\bibfnamefont {G.}~\bibnamefont {Romero}},
  \bibinfo {author} {\bibfnamefont {I.}~\bibnamefont {Lizuain}}, \bibinfo
  {author} {\bibfnamefont {J.~J.}\ \bibnamefont {Garc\'{\i}a-Ripoll}}, \ and\
  \bibinfo {author} {\bibfnamefont {E.}~\bibnamefont {Solano}},\ }\href
  {\doibase 10.1103/PhysRevLett.105.263603} {\bibfield  {journal} {\bibinfo
  {journal} {Phys. Rev. Lett.}\ }\textbf {\bibinfo {volume} {105}},\ \bibinfo
  {pages} {263603} (\bibinfo {year} {2010})}\BibitemShut {NoStop}%
\bibitem [{\citenamefont {Kieu}(2004)}]{PhysRevLett.93.140403}%
  \BibitemOpen
  \bibfield  {author} {\bibinfo {author} {\bibfnamefont {T.~D.}\ \bibnamefont
  {Kieu}},\ }\href {\doibase 10.1103/PhysRevLett.93.140403} {\bibfield
  {journal} {\bibinfo  {journal} {Phys. Rev. Lett.}\ }\textbf {\bibinfo
  {volume} {93}},\ \bibinfo {pages} {140403} (\bibinfo {year}
  {2004})}\BibitemShut {NoStop}%
\bibitem [{\citenamefont {Irish}(2007)}]{Irish2007}%
  \BibitemOpen
  \bibfield  {author} {\bibinfo {author} {\bibfnamefont {E.~K.}\ \bibnamefont
  {Irish}},\ }\href {\doibase 10.1103/PhysRevLett.99.173601} {\bibfield
  {journal} {\bibinfo  {journal} {Phys. Rev. Lett.}\ }\textbf {\bibinfo
  {volume} {99}},\ \bibinfo {pages} {173601} (\bibinfo {year}
  {2007})}\BibitemShut {NoStop}%
\bibitem [{\citenamefont {Yu}\ \emph {et~al.}(2012)\citenamefont {Yu},
  \citenamefont {Zhu}, \citenamefont {Liang}, \citenamefont {Chen},\ and\
  \citenamefont {Jia}}]{Yu2012analytical}%
  \BibitemOpen
  \bibfield  {author} {\bibinfo {author} {\bibfnamefont {L.}~\bibnamefont
  {Yu}}, \bibinfo {author} {\bibfnamefont {S.}~\bibnamefont {Zhu}}, \bibinfo
  {author} {\bibfnamefont {Q.}~\bibnamefont {Liang}}, \bibinfo {author}
  {\bibfnamefont {G.}~\bibnamefont {Chen}}, \ and\ \bibinfo {author}
  {\bibfnamefont {S.}~\bibnamefont {Jia}},\ }\href {\doibase
  10.1103/PhysRevA.86.015803} {\bibfield  {journal} {\bibinfo  {journal} {Phys.
  Rev. A}\ }\textbf {\bibinfo {volume} {86}},\ \bibinfo {pages} {015803}
  (\bibinfo {year} {2012})}\BibitemShut {NoStop}%
\bibitem [{\citenamefont {Datta}\ and\ \citenamefont
  {Gharibian}(2009)}]{PhysRevA.79.042325}%
  \BibitemOpen
  \bibfield  {author} {\bibinfo {author} {\bibfnamefont {A.}~\bibnamefont
  {Datta}}\ and\ \bibinfo {author} {\bibfnamefont {S.}~\bibnamefont
  {Gharibian}},\ }\href {\doibase 10.1103/PhysRevA.79.042325} {\bibfield
  {journal} {\bibinfo  {journal} {Phys. Rev. A}\ }\textbf {\bibinfo {volume}
  {79}},\ \bibinfo {pages} {042325} (\bibinfo {year} {2009})}\BibitemShut
  {NoStop}%
\bibitem [{\citenamefont {Allende}\ \emph {et~al.}(2015)\citenamefont
  {Allende}, \citenamefont {Altbir},\ and\ \citenamefont
  {Retamal}}]{Allende2015simulated}%
  \BibitemOpen
  \bibfield  {author} {\bibinfo {author} {\bibfnamefont {S.}~\bibnamefont
  {Allende}}, \bibinfo {author} {\bibfnamefont {D.}~\bibnamefont {Altbir}}, \
  and\ \bibinfo {author} {\bibfnamefont {J.~C.}\ \bibnamefont {Retamal}},\
  }\href {\doibase 10.1103/PhysRevA.92.022348} {\bibfield  {journal} {\bibinfo
  {journal} {Phys. Rev. A}\ }\textbf {\bibinfo {volume} {92}},\ \bibinfo
  {pages} {022348} (\bibinfo {year} {2015})}\BibitemShut {NoStop}%
\bibitem [{\citenamefont {Paauw}\ \emph {et~al.}(2009)\citenamefont {Paauw},
  \citenamefont {Fedorov}, \citenamefont {Harmans},\ and\ \citenamefont
  {Mooij}}]{Paauw2009tuning}%
  \BibitemOpen
  \bibfield  {author} {\bibinfo {author} {\bibfnamefont {F.~G.}\ \bibnamefont
  {Paauw}}, \bibinfo {author} {\bibfnamefont {A.}~\bibnamefont {Fedorov}},
  \bibinfo {author} {\bibfnamefont {C.~J. P.~M.}\ \bibnamefont {Harmans}}, \
  and\ \bibinfo {author} {\bibfnamefont {J.~E.}\ \bibnamefont {Mooij}},\ }\href
  {\doibase 10.1103/PhysRevLett.102.090501} {\bibfield  {journal} {\bibinfo
  {journal} {Phys. Rev. Lett.}\ }\textbf {\bibinfo {volume} {102}},\ \bibinfo
  {pages} {090501} (\bibinfo {year} {2009})}\BibitemShut {NoStop}%
\bibitem [{\citenamefont {Schwarz}\ \emph {et~al.}(2013)\citenamefont
  {Schwarz}, \citenamefont {Goetz}, \citenamefont {Jiang}, \citenamefont
  {Niemczyk}, \citenamefont {Deppe}, \citenamefont {Marx},\ and\ \citenamefont
  {Gross}}]{schwarz2013gradiometric}%
  \BibitemOpen
  \bibfield  {author} {\bibinfo {author} {\bibfnamefont {M.~J.}\ \bibnamefont
  {Schwarz}}, \bibinfo {author} {\bibfnamefont {J.}~\bibnamefont {Goetz}},
  \bibinfo {author} {\bibfnamefont {Z.}~\bibnamefont {Jiang}}, \bibinfo
  {author} {\bibfnamefont {T.}~\bibnamefont {Niemczyk}}, \bibinfo {author}
  {\bibfnamefont {F.}~\bibnamefont {Deppe}}, \bibinfo {author} {\bibfnamefont
  {A.}~\bibnamefont {Marx}}, \ and\ \bibinfo {author} {\bibfnamefont
  {R.}~\bibnamefont {Gross}},\ }\href
  {http://stacks.iop.org/1367-2630/15/i=4/a=045001} {\bibfield  {journal}
  {\bibinfo  {journal} {New Journal of Physics}\ }\textbf {\bibinfo {volume}
  {15}},\ \bibinfo {pages} {045001} (\bibinfo {year} {2013})}\BibitemShut
  {NoStop}%
\end{thebibliography}%
\bibliographystyle{apsrev4-1}
\appendix 
\section{Energy spectrum of the quantum Rabi model} \label{ap0}
\begin{figure}[t!] 
	\centering
	\includegraphics[width=1.1\linewidth]{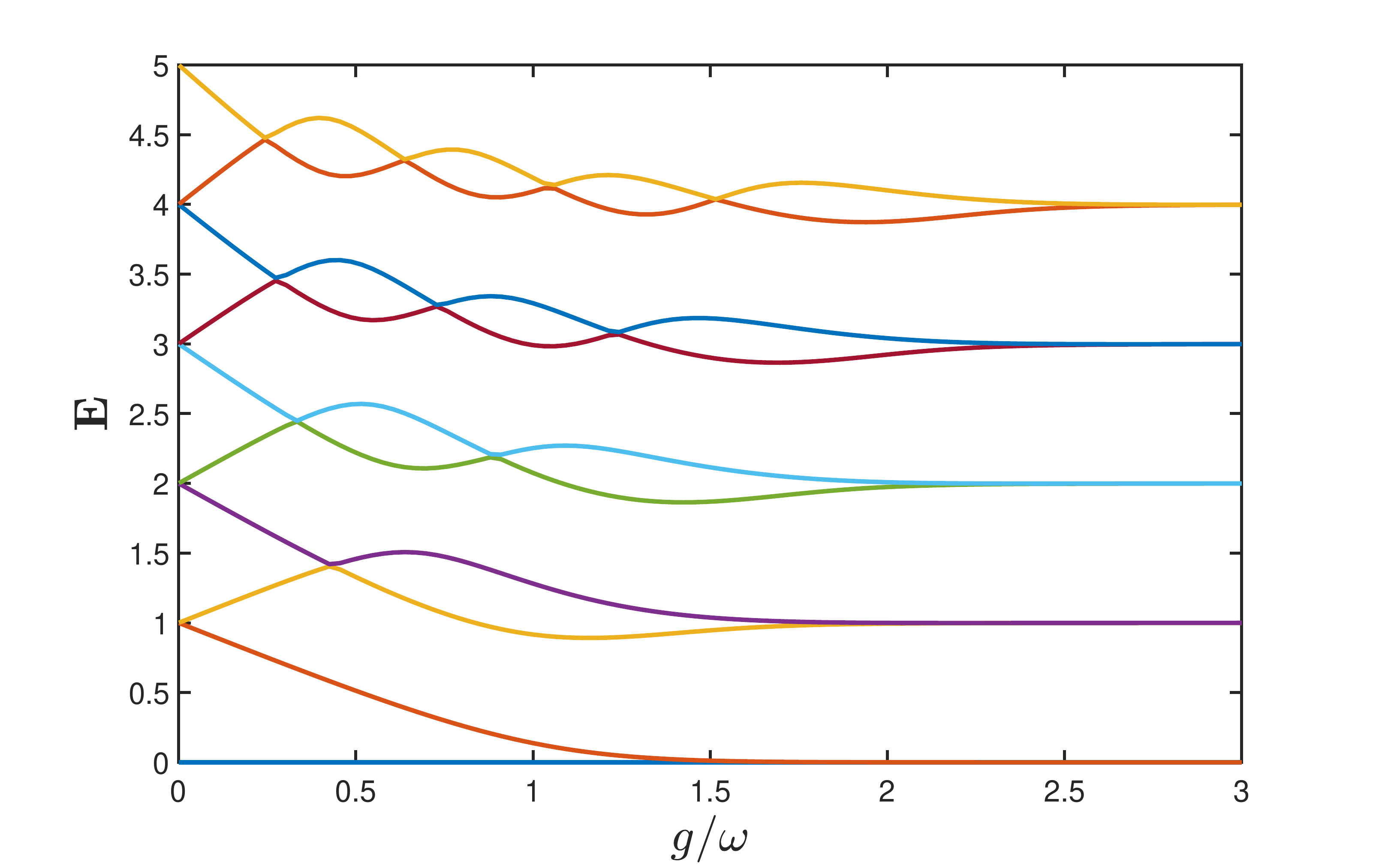}
	\caption{Energy spectrum of the quantum Rabi model relative to the ground state as a function of coupling parameter $g/\omega$.}
	\label{fig:Rabispectrum}
\end{figure}

The quantum Rabi model described by the Hamiltonian
\begin{equation}
H = \hbar \frac{\Omega}{2} \sigma_{z} + \hbar \omega a^{\dagger} a + \hbar g \sigma_{x} (a^{\dagger} + a),
\end{equation}
has been studied elsewhere. Here, we focus on the anharmonicity of its energy spectrum which is shown in Fig.\ref{fig:Rabispectrum} as a function of the ratio $g/\omega$. All the physics explained in this article is well captured by the ground and first excited state of the quantum Rabi model.
\section{Approximation for the energies of the quantum Rabi model} \label{ap1}
In order to compute the eigenenergies of the quantum Rabi model we follow Refs. \cite{Irish2007,Yu2012analytical}. In particular, we write the quantum Rabi model in a form which allows an expansion in terms of the creation and annihilation operators. This expression can be truncated to obtain a diagonizable Hamiltonian that leads to the approximated energy levels of the quantum Rabi model, this in turn will allow us to derive the positive work condition of Eq.(\ref{posw}).\par
 We proceed by rotating the two-level system such that the QRM can be written as:
\begin{equation}
H = \hbar \omega a^{\dagger} a - \hbar \frac{\Omega}{2} \sigma_{x} + \hbar g (a^{\dagger} + a) \sigma_{z}. 
\end{equation}
Next, we apply the displacement operator $D(g/\omega) = \exp(\frac{g}{\omega}\sigma(a^{\dagger} -a ))$ and obtain
\begin{equation}
	\begin{aligned}
H_{2} \equiv & DHD^{\dagger} =  \hbar \omega a^{\dagger} a - \hbar \frac{g^{2}}{2} \\
                             & - \hbar \frac{\Omega}{2} e^{\frac{-2g^{2}}{\omega^{2}}} \left(\begin{array}{cc} 0 & e^{\frac{2g}{\omega} a^{\dagger}}e^{\frac{-2g}{\omega} a^{\dagger}}\\ e^{\frac{-2g}{\omega} a^{\dagger}}e^{\frac{2g}{\omega} a^{\dagger}} & 0 \end{array}\right).
\end{aligned}
\end{equation}
Now expanding the exponential operators as
\begin{equation}
e^{\frac{2g}{\omega} a^{\dagger}}e^{\frac{-2g}{\omega} a^{\dagger}} = \sum_{n,m} \frac{(2g/\omega)^{n} a^{\dagger n}}{n!} (-1)^{m} \frac{(2g/\omega)^{m} a^{m}}{m!},  
\end{equation}
the Hamiltonian becomes: 
   \begin{figure}[t!] 
	\centering
	\includegraphics[width=1.1\linewidth]{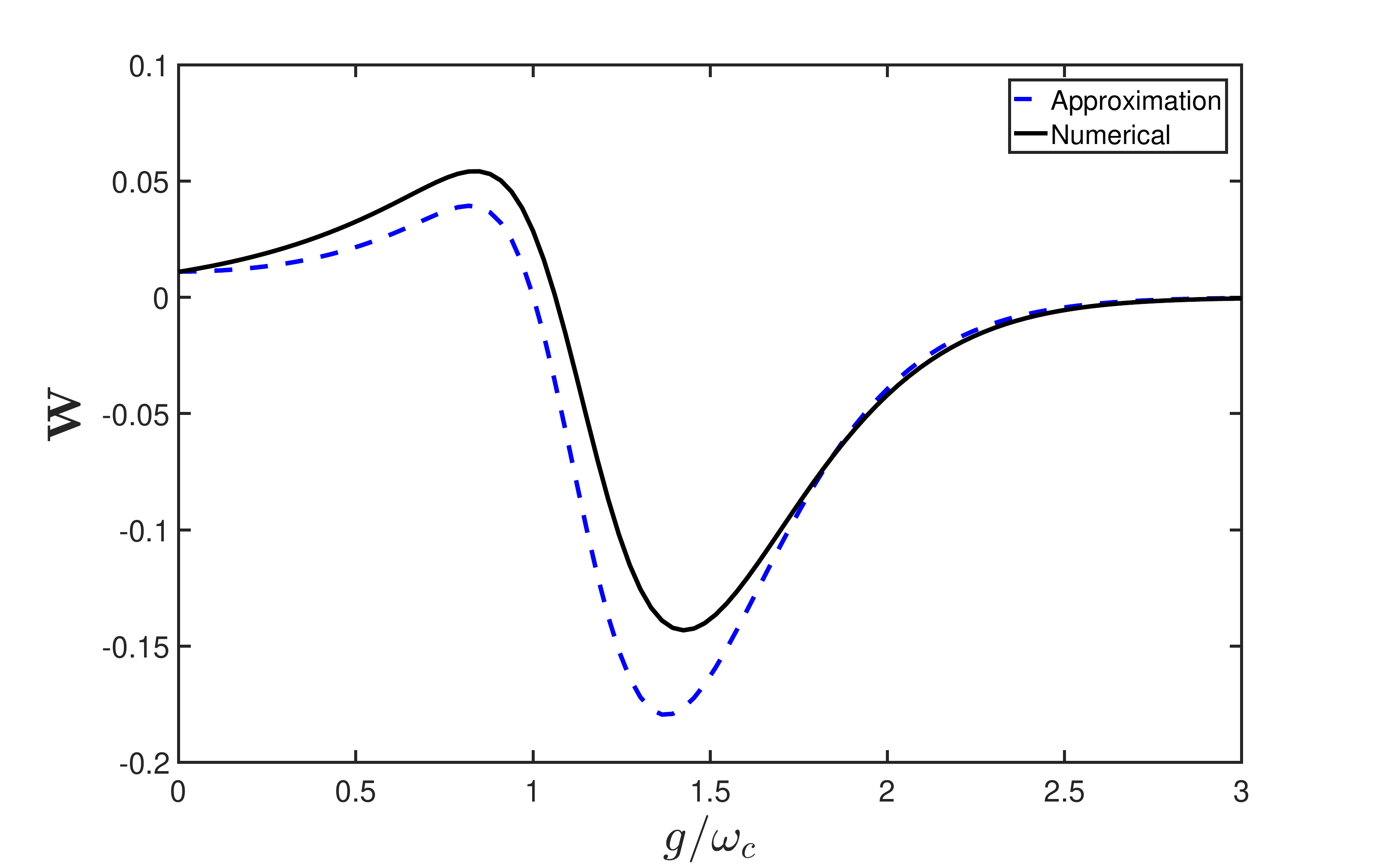}
	\caption{ Work contribution of the first excited state $W_{1}$ as obtained by the approximation presented (blue-dashed line) compared with the numerically-determined values (solid black line). Where we have used $T_{c} = 19$mK and $T_{h} = 9T_{c}$.}
	\label{fig:Work_aprox}
\end{figure} 
\begin{equation}
\begin{aligned}
H_{3} = & \hbar \omega a^{\dagger} a - \hbar \frac{g^2}{\omega} - \hbar \frac{\Omega}{2} e^{-2\frac{g^{2}}{\omega^{2}}} \Theta \sigma_{x} \\
    & - \hbar \frac{g}{\omega} \Omega  e^{-2\frac{g^{2}}{\omega^{2}}}  ( a^{\dagger} A - A a) ( \sigma_{+} - \sigma_{-}) + ...
\end{aligned}
\label{app2}
\end{equation} 
where 
\begin{eqnarray}
\Theta = \sum_{n=0} = \frac{(-1)^{n} (\frac{2g}{\omega})^{2n} }{(n!)^{2}} a^{\dagger n} a^{n} \nonumber \\ 
A = \sum_{n=0} = \frac{(-1)^{n} (\frac{2g}{\omega})^{2n + 1} }{n!(n+1)!} a^{\dagger n} a^{n}.
\end{eqnarray}
We recognize the last term of Eq.(\ref{app2}) as an interaction term which is exponentially damped by a factor $e^{-2\frac{g^{2}}{\omega^{2}}}$, this justifies the application of the RWA and leads to a diagonalization of $H_{3}$ in a similar way as for the Jaynes-Cummings Hamiltonian. However, we seek a simple expression for the approximated energies of $H$, thus we shall only consider the free term of Eq.(\ref{app2}), that is:
\begin{equation}
H_{4} = \hbar \omega a^{\dagger} a - \hbar \frac{g^2}{\omega} - \hbar \frac{\Omega}{2} e^{-2\frac{g^{2}}{\omega^{2}}} \Theta \sigma_{x}.
\end{equation}
Now, the eigenstates of the above Hamiltonian are of the form  $|\pm,n \rangle $, with energies: 
\begin{equation} 
E_{\pm,n} = -\hbar \omega n - \hbar \frac{g^2}{\omega} \mp \hbar \frac{\Omega}{2} e^{-2\frac{g^{2}}{\omega^{2}}} L_{n}(4\frac{g^2}{\omega^2}),
\end{equation}
where $L_{n}(4\frac{g^2}{\omega^2}) $ is a Laguerre polynomial of order $n$. Then the energy of the ground and first excited states are:
\begin{eqnarray}
E_{G} =  - \hbar \frac{g^2}{\omega} - \hbar \frac{\Omega}{2} e^{-2\frac{g^{2}}{\omega^{2}}}\\
E_{1} =  - \hbar \frac{g^2}{\omega} + \hbar \frac{\Omega}{2} e^{-2\frac{g^{2}}{\omega^{2}}}.
\end{eqnarray}
Now we can obtain an approximated expression for the work contribution of the first excited state $W_{1}$, which dominates the behavior of the harvested work. We choose $\omega_{h} = \Omega_{h} = R \omega$ and   $\omega_{c} = \Omega_{c} = \omega$, and write the inverse temperatures for the cold and hot reservoir as $\beta_{h} = 1/k_{B}T_{h}$ and $\beta_{c} = 1/k_{B}T_{c}$. Then, for a two-level approximation, Eq.(\ref{Work}) of the main text reads
\begin{equation}
W = (b_{h} - b_{c}) (\tanh(\beta_{c} b_{c}) - \tanh(\beta_{h} b_{h})),
\label{apwork}
\end{equation}
where  $b_{h} = \hbar \frac{R \omega}{2} e^{-2\frac{g^{2}}{R^{2}\omega^{2}}}$, $b_{c} = \hbar \frac{\omega}{2} e^{-2\frac{g^{2}}{\omega^{2}}}$ and $R=\frac{\omega_{h}}{\omega_{c}} > 1$. \par 
Figure \ref{fig:Work_aprox} shows $W_{1}$ as calculated from Eq.(\ref{apwork}) compared with the numerical calculation. As can be seen, while the difference between the approximation and the numerical calculation cannot be neglected, it is clear that Eq.(\ref{apwork}) captures the behavior of $W_{1}$. Now, from Eq.(\ref{apwork}) we can obtain the condition for the positive work regime by imposing $W>0$, which leads to 
\begin{equation}
\frac{T_{h}}{T_{c}} > \frac{\omega_{h}}{\omega_{c}} e^{2\frac{g^2}{\omega^2} \big( 1 - \frac{1}{R^2} \big) }.
\end{equation}
We can also express the above equation as follows
\begin{equation}
\frac{g}{\omega} < \sqrt{ \frac{1}{2} \frac{R^{2}}{R^{2} - 1}  \textnormal{ln}\bigg(\frac{1}{R} \frac{T_{h}}{T_{c}}\bigg)}.
\label{workcond}
\end{equation}
\textcolor{white}{We have shown how the anharmonicity and degeneracy of the spectrum of the quantum Rabi model give rise to an operation regime transition from heat engine into refrigerator as the light-matter coupling increases. By considering a generalized quantum Rabi model we found that  the quantum correlations reduction in the hot bath stage  and in the compression stage of the cycle can lead to enhanced positive work extraction and efficiency respectively. Finally we have shown that an easier implementation of the model by considering only a change in the qubit frequency during the adiabatic stages still maintains the relation we have shown between quantum correlations and positive work extraction.}\par

\end{document}